%% file: SLR4Arxiv.tex
\documentclass{article}

\usepackage{arxiv}

\usepackage[utf8]{inputenc} 
\usepackage[T1]{fontenc}    
\usepackage{hyperref}       
\usepackage{url}            
\usepackage{booktabs}       
\usepackage{amsfonts}       
\usepackage{nicefrac}       
\usepackage{microtype}      
\usepackage{graphicx}
\usepackage[numbers]{natbib}
\usepackage{doi}

\input{packages}
\input{commands}
\input{acronyms}
\usepackage{hyperref} 
\usepackage{cleveref}
\usepackage{orcidlink}

\graphicspath{{figs/}{figures/}{pictures/}{images/}{./}} 

\title{Biomedical Image Segmentation: A Systematic Literature Review of Deep Learning Based Object Detection Methods}

\author{ Fazli Wahid$^{1,2}$\orcidlink{0000-0003-4091-6817}, Yingliang Ma$^{2}$*\orcidlink{0000-0001-5770-5843}, Dawar Khan$^{3}*\orcidlink{0000-0001-5864-1888}$, Muhammad Aamir$^{1}$\orcidlink{0000-0002-4999-7740}, and Syed U. K. Bukhari$^{4}$ 
    \\
    $^{1}$ College of Science and Engineering, School of Computing, University of Derby, United Kingdom.\\ E-mail: f.wahid@derby.ac.uk \\
    $^{2}$ School of Computing Sciences, University of East Anglia,  United Kingdom.\\ 
       $^{3}$King Abdullah University of Science and Technology (KAUST), Saudi Arabia.\\
        E-mail: dawar.khan@kaust.edu.sa. \\ 
        $^{4}$ Idrak AI Ltd, United Kingdom (Email: drusama@idrakai.com).\\
        $^{*}$ Corresponding authors: yingliang.ma@uea.ac.uk; dawar.khan@kaust.edu.sa
}
 
\date{}

\renewcommand{\shorttitle}{Biomedical Image Segmentation: A Systematic Literature Review }

\begin{document}
\maketitle

%
\begin{figure*}[!h]
    \centering \vskip -1.5cm
    \includegraphics[width=\linewidth]{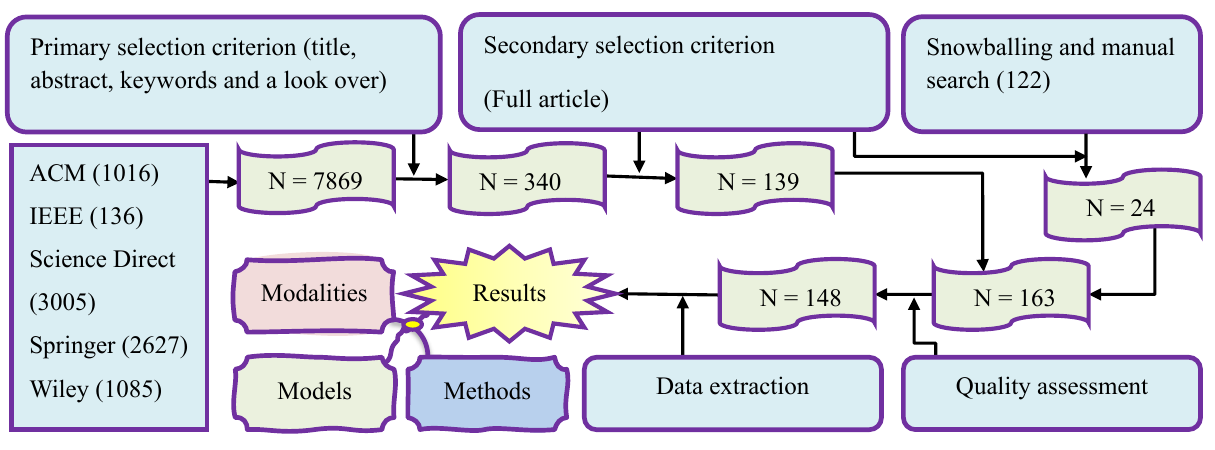}
    \vskip -0.25cm
    \caption{Overview of the Research Methodology and Search Results: A step-by-step flow from article search to final selection and quality assessment, leading to the results. }
    \label{fig:segMethods}
\end{figure*} 

\begin{abstract}
\input{content/00-Abstract.tex} 
\end{abstract}


\maketitle 

\input{content/01-introduction}

\input{content/02-ResearchMethod}
\input{content/03-Modalities}
\input{content/04-ObjDetections}

\input{content/05-Results}
\input{content/06-FutureDirections}

\input{content/07-Conclusion} 
\section*{Acknowledgments}%
This work was supported by the Engineering and Physical Sciences Research Council (grant number: EP/X023826/1).
\FloatBarrier

 \bibliographystyle{unsrtnat}

\bibliography{SLR4Arxiv}
\end{document}

%% file: packages.tex
\usepackage{graphicx}
\usepackage{arxiv}
\usepackage{acro}

\usepackage[utf8]{inputenc} 
\usepackage[T1]{fontenc}    
\usepackage{setspace}
\usepackage{hyperref}       
\usepackage{url}            
\usepackage{booktabs}       
\usepackage{amsfonts}       
\usepackage{nicefrac}       
\usepackage{microtype}      
\usepackage{lipsum}         
\usepackage{natbib}
\usepackage{doi}

\usepackage{subcaption}

\usepackage{algorithm} 
\usepackage{algorithmic}  
\usepackage[algo2e]{algorithm2e}

\usepackage[acronym]{glossaries}
\usepackage{booktabs}

\usepackage{textcomp}
\usepackage{xspace}

\usepackage{amsmath} 
\usepackage{setspace}

\usepackage{placeins} 

\usepackage{enumitem}
\usepackage{balance}
\usepackage{lineno}

%% file: commands.tex
\newcommand{\h}[1]{}

%% file: acronyms.tex
\DeclareAcronym{cvt}{
  short=CVT,
  long=centroidal {V}oronoi tessellation,
}
\DeclareAcronym{Cryo-em}{
  short=cryo-EM,
  long=cryogenic electron microscopy ,
} 
\DeclareAcronym{Cryo-et}{
  short=cryo-ET,
  long=cryogenic electron tomography,
} 
\DeclareAcronym{sssr}{
  short=SSSR,
  long=simple and scalable surface reconstruction,
}
\DeclareAcronym{mat}{
  short=MAT,
  long=medial axis transform,
}
\DeclareAcronym{SDF}{
  short=SDF,
  long=Signed Distance Field,
}
\DeclareAcronym{FIB}{
  short=FIB,
  long=focused ion beam,
}
\DeclareAcronym{SEM}{
  short=SEM,
  long=scanning electron microscopy
}
\DeclareAcronym{FIB-SEM}{
  short=\acs{FIB}-\acs{SEM},
  long=\acl{FIB}-\acl{SEM}
}
\DeclareAcronym{PSR}{
  short=PSR,
  long=Poisson Surface Reconstruction,
}
\DeclareAcronym{MD}{
  short=MD,
  long=molecular dynamics,
}

%% file: content/00-Abstract.tex
 Biomedical image segmentation plays a vital role in  diagnosis of diseases across various organs. Deep learning-based object detection methods are commonly used for such  segmentation. There exists an extensive research in this topic. However, there is no standard review on this topic. Existing surveys often lack a standardized approach or focus on broader segmentation techniques.
In this paper, we conducted a systematic literature review (SLR), collected and analysed 148 articles  that explore deep learning object detection methods for biomedical image segmentation. We critically analyzed these methods, identified the key challenges, and discussed the future directions. From the selected articles we extracted the results including the deep learning models, targeted imaging modalities, targeted diseases, and the metrics for the analysis of the methods. The results have been presented  in tabular forms. The results are presented in three major categories including two stage detection models, one stage detection models and point-based detection models.  Each article is individually analyzed along with its pros and cons.  Finally, we discuss open challenges, benefits, and future directions. This SLR aims to provide the research community with a quick yet deeper understanding of these segmentation models, ultimately facilitating the development of more powerful solutions for biomedical image analysis.

%% file: content/01-Introduction.tex
\section{Introduction}
\label{sec:introduction}  
Biomedical image segmentation plays a vital role in the analysis and diagnosis of diseases related to various organs. In this regard, the accurate identification of the region of interest in the biomedical images is a significant phase for identifying affected regions in the image~\cite{Wang2019DeepIGeoS}. There are several alternative methods to achieve this segmentation.  object detection based methods is a type of instance segmentation that is very common in this domain. Deep learning-based object detection methods~\cite{CalderonRamirez2022survey,Minaee2022survey}  offer significant advantages over the other alternatives. Considering the prominence of deep learning based object detection-based methods for biomedical image segmentation, we believe a comprehensive review of this field can be a a valuable work to provide a quick insight into the field. Although, there are few reviews (\cref{sec:rs}) in the related topic, there is no standard review on the target direction.   \\
Inspired by the systematic literature review (SLR) methodology (\cref{sec:Methodology}) from software engineering domain~\cite{Kitchenham07guidelinesfor}, this paper presents a comprehensive SLR of the object detection based deep learning methods used for biomedical image segmentation.  Following SLR guidelines, we collected 150 articles (primary studies) in an unbiased manner. After collecting the primary studies, we extracted the results. The results include the deep learning models, modalities, a list of diseases, and the metrics used for the analysis of the segmentation methods. We also provide critical analysis of the methods and presented each of the article with their pros and cons. 
\subsection{Background and Motivation}  
 Segmentation of biomedical images can be categorized into two main categories including semantic segmentation and instance segmentation. In semantic segmentation, each pixel of an image is analyzed and classified into a particular class. This type of segmentation particularly helps in  capturing the quantitative shapes of objects (in the case of biomedical imaging, a disease or affected region) in the image~\cite{taghanaki2021deep}. The deep learning models applied for semantic segmentation are typically found in three  categories including region-based (RB) segmentation~\cite{millioni2010analysis}, fully convolutional neural network (FCN)~\cite{roth2018deep,buda2019association,zhu2021coronary}, and weakly supervised or semi-supervised learning (SSL) approaches~\cite{kim2016evaluation,isin2016review}.\\
In the instance segmentation, each pixel of image is assigned to some specific instance which has a significant role in biomedical imaging for the identification and analysis of various types of diseases or same type of disease with different locations in the image. Different types of algorithms are used for instance segmentation e.g., supervised learning (SL) models and semi-supervised learning (SSL) approaches. Each approach has its own drawbacks and benefits associated with them. Recently, fully supervised models have got proper attention for biomedical image instance segmentation which are based on human experts’ knowledge for annotations. These models are very laborious and expensive in terms of obtaining dense and accurate annotations ~\cite{lee2021learning}. To overcome the problem of time consumption and human efforts associated with annotation for supervised learning approaches, some semi-supervised models have been proposed which require small set of annotated biomedical image samples as compared to the fully supervised instance segmentation models ~\cite{huang2022semi}. 
The instance segmentation is mainly categorized into two main types namely proposal-free ~\cite{liu2023soma} and proposal-based instance segmentation ~\cite{liu2021panoptic}. The former category is based on morphological properties and instance-aware features whereas the latter is type is based on object detection approach which is further divided into two stage detectors, one stage detectors, and point-based detectors or advanced detectors. \\ 
 Biomedical image segmentation has gained a significant attention from researchers, resulting in the development of various methods across different categories. Among these, object detection-based methods utilizing deep learning have emerged as particularly effective for biomedical image segmentation. However, unlike primary studies, currently there is no comprehensive review of this specific topic (see~\autoref{sec:rs}). To the best of our knowledge, there is no systematic literature review (SLR) on the topic. There are few review articles but they are not following some standard methodologies. Therefore, an SLR is essential to consolidate existing knowledge, identify research gaps, and provide a cohesive understanding of the advancements in this field. This review aims to bridge that gap by thoroughly examining and synthesizing the literature on deep learning-based object detection methods for biomedical image segmentation.

 \begin{table*}[!htbp]
  \centering
  \caption{Related Surveys and Comparison with Our SLR.}
  \resizebox{\linewidth}{!}{ 
    \begin{tabular}{|c|p{1cm}|p{1cm}|p{5.0cm}|p{3.5cm}|p{6cm}|p{5.65cm}|}
      \hline
      \textbf{S.No} & \textbf{Review ID} & \textbf{Covered up to Year} & \textbf{Imaging Modalities} & \textbf{Processing Tasks Performed} & \textbf{Models Targeted} & \textbf{Diseases/Organs Covered} \\
      \hline
      1 & R01~\cite{isin2016review} & 2016 & MRI (T1, T1-Gd, T2, FLAIR) & Automatic, Semantic Segmentation & CNN, ML, Generative Models & Brain Tumor \\
      \hline
      2 & R02~\cite{anwar2018medical} & 2018 & CT, MRI, PET, X-Rays, Ultrasound & Segmentation, Classification, Diagnosis & CNN different variants + ML models & Lungs, Breast, Brain, Thyroid \\
      \hline
      3 & R03~\cite{haque2020deep} & 2019 & Clinical, X-Rays, CT, MRI, Ultrasound, OCT, Microscopic Images & Semantic Segmentation, Classification, Analysis & ML, CNN, RBM, Auto encoders, GANs, RNN, FCN, U-Net, ResNet, V-Net & Brain, Breast, Liver, Spleen, Heart, Stomach, Ventricles, Arteries, Skin \\
      \hline
      4 & R04~\cite{alzahrani2021biomedical} & 2019 & PET, CT, X-Ray, Ultrasound, Histological Images & Classification, Segmentation & VGG-16, GoogleNet, AlexNet, ResNet, ResUnet, SegNet, UNet++, RD-Unet, RNN, ML Models & Not Mentioned \\
      \hline
      5 & R05~\cite{zhou2019review} & 2019 & T1, T2, T1C, FLAIR, T1-DUAL, T2-SPIR, T1-IR, CT, PET, MRI & Semantic Segmentation & LeNet, ZFNet, GoogleNet, DenseNet, ResNet, AlexNet, U-Net, RNN, SVM, FCN, CRF & Brain Tumor, Brain Stroke, Ischemic stroke lesion, Abdominal Organs, Intervertebral Disc \\
      \hline
      6 & R06~\cite{kar2021review} & 2020 & Mammograms, MRI, Endoscopic OCT, Microscopic images, Radiographs, Colonoscopy, CT, X-Rays, Fundus Camera, Multi-modal, 3D Images & Semantic Segmentation & U-Net++, Dual U-Net, SegNet, PSPNet, U-Net, ReNet, ResNet, VGG-16, AlexNet, GoogleNet, GAN, DBN, RBM, AE, SAE, RNN, FCN & Breast, Brain, Eye, Pancreas, Gland, Chest, Head, Tissues, Lungs, Cells \\
      \hline
      7 &R07~\cite{taghanaki2021deep}& 2020 & Microscopy, MRI, CT, EM, Fundus, X-Ray, PET & Semantic Segmentation & SegNet, FCN, Auto-DeepLab, RefineNet, ResNet-38, DeepLabV3 & Prostate, Lung Nodule, Pancreas, Abdomen, Spleen, Breast, Heart, Spinal Cord, Retinal Vessel, Brain, Colon, Vertebral Body \\
      \hline
      8 & R08~\cite{tajbakhsh2020embracing} & 2020 & MRI, CT, X-Rays, LGE, bSSFP, DRR, Ultrasound, Angiography, FLAIR, Fundus images, Histography & Semantic Segmentation, Classification, Detection & Self-Learning Models, Semi-Supervised Models, Unsupervised Models, Ensembled Models & Lungs, Abdomen, Breast, Tumor, Brain, Pancreas, Kidney, Liver, Heart \\
      \hline
      9 & R09~\cite{iqbal2022generative} & 2021 & CT, MRI, X-Ray, Mammography, Ultrasound & Segmentation & Vanilla-GAN, Cycle-GAN, Patch-GAN, Style-GAN, DC-GAN & Retina, Brain, Skin, Heart, Breast, Bone, Lungs \\
      \hline
      10 & R10~\cite{punn2022modality} & 2021 & Ultrasound, PET, X-Ray, CT, FLAIR & Segmentation & U-Nets’ Variants (Better U-Nets, Inception U-Nets, Attention U-Nets, Ensemble U-Nets) & Heart, Lungs, Chest, Bones, Blood Vessels, Soft Tissues, Skin \\
      \hline
      11 & R11~\cite{qureshi2023medical} & 2021-2022 & CT, MRI, US, Microscopy, Histology, EM, PET, Colonoscopy & Semantic Segmentation & FCN, U-Net, Gamma-Net, SSIM, AFPNet, ResNet & Brain, Chest, Breast, Skin, Prostate \\
      \hline
      12 & R12~\cite{fernando2022deep} & 2022 & 3D-MRI & Semantic Segmentation & Statistical Methods, Deep Learning Models, Hybrid Models & Brain \\
      \hline
      13 & R13~\cite{xiao2023transformers} & 2023 & MRI, CT & Semantic Segmentation & Different types of transformer models & Abdomen, Heart, Brain, Lungs \\
      \hline
      14 & R14~\cite{wang2022medical} & 2022 & X-Rays, CT, MRI, Ultrasound & Semantic Segmentation & U-Net, 3D-Net, RNN, ResPath, FED-Net, FCN, CDNN, GAN, Transformer models & Liver, Pancreas, Heart, Lungs, Brain, Kidney, Spleen, Skin, Thyroid, Prostate, Colon \\
      \hline
      15 & R15~\cite{iqbal2023survey} & 2023 & CT, MRI, X-Rays, Echocardiograms & Semantic Segmentation & CNN, U-Net, DeepLabV3, ResNet-18, RF, NB, DT, LR, KNN, XGBoost & Chest, Arteries, Heart \\
      \hline
      16 &R16~\cite{ilesanmi2023systematic} & 2023 & Fundus & Classification, Segmentation & CNN, FCNN, VGG, U-Net, LeNet-5, M-Net & Arteries, Veins, Retina \\
      \hline
      17 & Ours& 2024 & 30 different modalities & Object Detection Based segmentation & Two-stage detectors, One-stage detectors, Advanced detectors & Maximum Body Organs Covered \\ 
      \hline
    \end{tabular}
  }
  \label{tabII:relatedSurveys}
\end{table*}
\subsection{Related Surveys}
\label{sec:rs}
\Cref{tabII:relatedSurveys} presents related surveys in the field and their comparison with our current SLR. Many surveys have been conducted related to different types of segmentation methods of biomedical images.   All of them cover different types of biomedical image segmentation approaches with keeping in consideration various parameters in their scopes.  
As shown in \Cref{tabII:relatedSurveys}, the major attractions of researchers in the last decade have been semantic segmentation using different types of machine learning models, semi-supervised learning-based models, and supervised learning-based models. A very less and insufficient attention has been paid to instance segmentation with further low attention towards object detection-based segmentation models. ~\Cref{tabII:relatedSurveys} presents a comparison of all the existing surveys and distinction of our current survey. The main focus of our survey paper will be different types of object detection-based segmentation models applied for segmentation of biomedical imaging which will be the first survey conducted on this worth-attraction and important direction.  Furthermore, it is evident that no proper attention has been given to the segmentation of biomedical imaging based on object detection models which are very powerful and recent techniques for the diagnosis of medical images. One similar survey has been conducted by Gui~\cite{Gui2020RSMDPI}; however, their focus is on remote sensing, and they did not follow the Systematic Literature Review (SLR) methodology. \\ 
The lack of following the SLR or other standard methodologies has been consistent across all existing review articles in the field. This review will be among the pioneer articles in AI-based image processing, particularly in image segmentation.   SLR provides a structured predefined set of rules and comprehensive synthesis of research findings. It offers critical insights on a given topi, and thereby yields a well structured survey. It is unbiased as every step from searching to the final results presentation are clearly defined in an organized way.
\subsection{Contributions }
The key contributions include:
\begin{itemize}
\item We conducted the first SLR on this topic, where we selected 148 primary studies as our final list and conducted a comprehensive analysis.
\item Various types of biomedical imaging modalities extracted from the literature are presented along with their pros and cons  for their usage in the disease detection and diagnosis.
\item Deep learning models based on object detection algorithms used for biomedical image segmentation are extracted from the selected articles.   
\item We identified various challenges of object detection-based segmentation methods for biomedical images, and listed few directions for future research with a sketch of potential solutions.
\end{itemize}

%% file: content/02-ResearchMethod.tex
\section{Research Methodology}
\label{sec:Methodology} 
Inspired by   its significance, and quality, we followed the SLR guidelines~\cite{Kitchenham07guidelinesfor} to conduct this review. SLR provides a predefined rules and thereby provides an unbiased framework for  conducting a high-quality literature review. This methodology formulates the whole architecture of the review, providing step by step protocols. Our research methodology is further described in the following subsections. 
\subsection{Research Questions}
Research questions are the key fillers of the research providing a baseline for analysis the quality and content of the research. This SLR is based on the following research questions (all in the domain of deep learning for objected detection biomedical image segmentation). Each question yields a list of outcomes with critical analysis among them.
\begin{enumerate}[leftmargin=1.05cm, label={RQ \arabic*:}]
  \item What are the different imaging modalities used for detection and diagnosis of human diseases? 
  \item Which object-detection based deep learning models are applied for the segmentation of different biomedical imaging modalities?
  \item What are the major problems and issues associated with each object detection based method when applied for biomedical image segmentation?
  \item What types of human diseases have been detected in biomedical imaging using different segmentation models?
   \item What  are the statistical metrics used evaluation of the segmentation methods, and models used in the domain? 
\end{enumerate}
\subsection{Search Strategy}
The search was carried out with the following search string. 

\fbox{
  \begin{minipage}{42.85em}
(Biomedical OR Medical) AND Segmentation AND (``Deep Learning'' OR ``Machine Learning'') AND (``Object Detection'') AND (Disease OR Disorder OR Abnormality)
  \end{minipage}
}
\\ $\;$\\

\noindent We searched the articles using the mentioned serach string in the following libraries: 
\begin{enumerate}
\item ACM digital library~\href{https://dl.acm.org}{\texttt{acm.org}},
\item IEEE Xplore~\href{http://ieeexplore.ieee.org}{\texttt{ieee.org}},
\item SpringerLink~\href{https://link.springer.com}{\texttt{link.springer.com}},
\item ScienceDirect~\href{https://www.sciencedirect.com/}{\texttt{sciencedirect.com}}, and 
\item Wiley~\href{https://onlinelibrary.wiley.com/}{\texttt{onlinelibrary.wiley.com}}.
\end{enumerate}

 In addition, we identified some article via manual search from other venues, such as \textit{Eurographics digital library}~\href{http://diglib.eg.org/}{\texttt{diglib.eg.org/}} and  ~\href{https://arxiv.org/}{\texttt{arxiv.org}}. 

 All these databases have a vast diversity in the number of publications both from scientific field as well as clinical field.  The first search of relative literature was performed in November 2023.  
\subsection{Article's Selection Criteria }
The articles were selected based in two stages. The primary selection was based on the title, abstract and the related keywords. All the articles were reviewed by authors for finding the relevance of inclusion or exclusion according to criteria specified. In this stage, we selected 340 research articles. 
The final selection is based on the reading the full paper. In this stage a total of  139 (out of 340) articles were selected. In addition, 24 articles were selected through snowballing and manual search, making a total of 163 articles, which were then passed through quality assessment in the next step.   The articles were included/included on the following criteria.
\begin{itemize}
    \item We included peer-reviewed journal and conference papers, as well as non-peer-reviewed preprints, but excluded non-peer-reviewed reports and books. 
    \item We included the articles with major focus on object detection-based segmentation methods and excluded those related with general segmentation approaches and methods related specifically to clinical systems. 
    \item Previously conducted surveys and reviews were also excluded from the study.
 \item Research articles related to approaches of semantic segmentation and methods based on proposal-free were also excluded. 
\item Due to rapid development in the field of deep learning models and automatic segmentation methods in the last one decade, it is believed that including older articles will not be so fruitful and necessary. Resultantly, the research articles published after 2013 have been included and the older articles were excluded. 
\item We excluded non-English articles. 
\item We excluded articles on quality assessment. 
\end{itemize}
\subsection{Article's Quality Assessment}
\label{Sec:qty:ass} 
In addition, to the above selection criteria, we excluded 15 articles based on their quality. To evaluate the quality and relevance of each article for our study, we used a quantitative metric indicating articles quality. The metric was calucalted for each article as:

\begin{enumerate}
\item Citation Frequency: Number of citations per year on Google Scholar.
\begin{itemize}
\item Good (1): 8+ citations per year (or articles published after 2020)
\item Average (0.5): 5+ citations per year
\item Poor (0): Less than 5 citations per year
\end{itemize}
\item Publication Venue: Quality and relevance of the journal or conference.
\begin{itemize}
\item Good (1): Published in a high-quality, relevant journal or conference
\item Average (0.5): Published in a multidisciplinary venue
\item Poor (0): Published in a low-quality or irrelevant venue
\end{itemize}
\item Methodology: Novelty and article presentations.
\begin{itemize}
\item Good (1): Novel and well written/presented. 
\item Average (0.5): Moderately novel and/or Moderately presented
\item Poor (0): Lower novelty and/or poor writing/presentation 
\end{itemize}
\item Results Analysis: Accuracy and clarity of results analysis.
\begin{itemize}
\item Good (1): Thorough and accurate analysis with clear comparisons or theoretical proofs
\item Average (0.5): Moderate analysis with some clarity issues
\item Poor (0): Inadequate analysis
\end{itemize}
\end{enumerate}

Each question was rated with either 1, 0.5 or 0; and then the accumulative score was used as quality metric. There were 15 articles scoring the accumulative value below 2  and they were excluded from our selection.  
\subsection{Extracted Information from the Papers}
Information strictly related to object detection based biomedical imaging from each paper was extracted which includes following specifications.
\begin{itemize}
    \item  Any pre-processing method applied for enhancing the quality of biomedical images before actual processing by the relative models.
\item  The overall methodology adopted for automatic segmentation of biomedical images.
\item  The implementation details including programming languages, software tools, applications, and hardware specifications.
\item  Dataset’s description including total size, training and testing ratios, and cross validation if available.
\item  Performance evaluation metrics
\item  Comparative analysis with other models if they have performed and available.
\end{itemize}
Different manuscripts have performed their performance evaluation using different parameters but few of these standard parameters include accuracy, True Positive Rate (TPR, also called sensitivity), True Negative Rate (TNR, also called specificity), False Discovery Rate (FDR), Jaccard Index (JI), Dice Similarity Coefficient (DSC), Hausdorff Distance (HD), Intersection Over Union (IOU), area under receiver operating characteristic curve, and Kappa. Except the HD, the higher value of the metric indicates better performance and lower value indicates worse performance. In case of HD, lower value represents better performance and higher value shows worse performance ~\cite{mittal2022comprehensive}. 

\subsection{Presentation of the Findings}
 The information collected from the selected articles are categories into three categories.  First, the imaging modalities and the deep learning models identified in the literature are presented in section~\ref{sec:modalities} and section~\ref{sec:DLModels}, respectively. Then, section~\ref{sec:mainSegementation} presents all the segmentation methods in detail, discussing their pros and cons, and summarizes various findings in tabular form. Section~\ref{sec:futureDir} highlights various research challenges and research gaps for future research. Finally, the paper is concluded in section~\ref{Sec:conc}.

%% file: content/03-Modalities.tex
 
\section{BIOMEDICAL IMAGING MODALITIES}
\label{sec:modalities}
\begin{figure}[!htbp]
    \centering 
    \includegraphics[width=\linewidth]{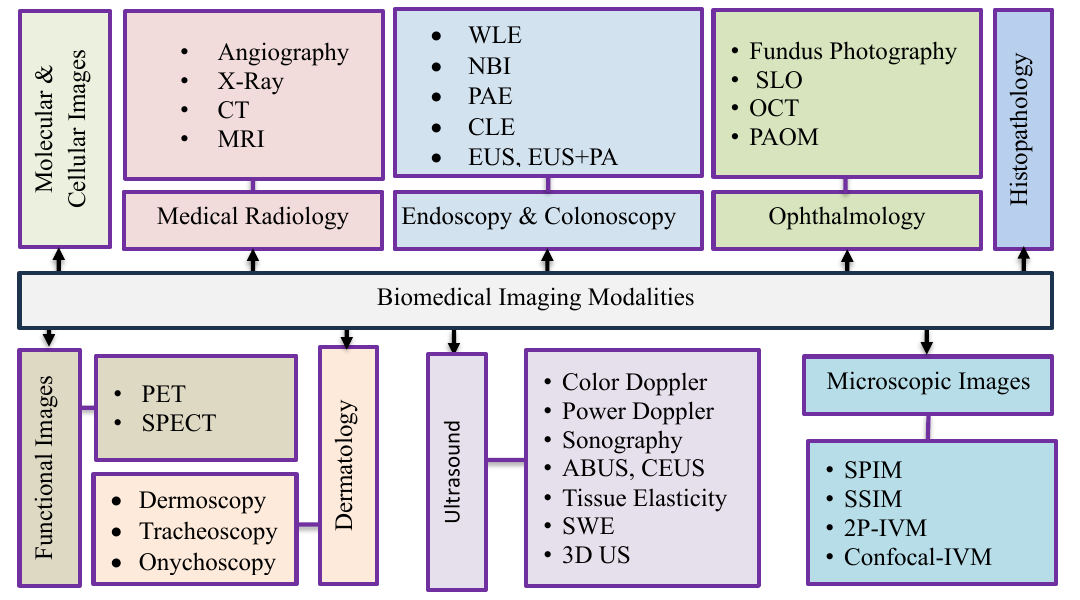}
    \caption{Biomedical Imaging Modalities}
    \label{fig:Modalities}
\end{figure}

There are many types of biomedical images’ modalities applied for detection, diagnosis, and treatment of various types of diseases. Depending on the capturing mechanisms, materials and equipment used, and the procedure followed for their diagnosis, these images have been categorized in different classes which are all covered in this section. Various types of biomedical imaging modalities are shown in~\Cref{fig:Modalities}, and described below.  \\
 \textbf{1) Medical Radiology Images:}
Although, medical radiology is a major branch of medical sciences that deals with detection, diagnosis, and treatment of various types of diseases using imaging technology, based on some standardized characterization of few imaging modalities, they have been classified under this category. The important medical radiology imaging modalities include X-rays, Computer Tomography (CT), Magnetic Resonance Imaging (MRI), and angiography images which are special types of X-rays or CT images ~\cite{smith2010introduction}.\\
\textbf{2) Ultrasound Images (US):}
The second important category of medical images modalities is ultrasound images which are used for diagnosis of various organs e.g., abdomen, kidneys, breast, spleen, liver, pancreas, and lungs. The important modalities of ultrasound are sonography images, color doppler, automatic breast ultrasound (ABUS), contrast enhanced ultrasound (CEUS), tissue elasticity, shear wave elastography, 3D ultrasound, and power doppler ~\cite{iranmakani2020review}.\\
\textbf{3) Endoscopy and Colonoscopy Images:}
The colonoscopy and endoscopy images are used for the diagnosis and treatment of gastrointestinal tract which starts from mouth, goes through pharynx and oesophagus, stomach and small intestine, large intestine and colon. The important categories of these images include white light endoscopy (WLE), narrowband imaging (NBI), optical coherence tomography (OCT), confocal laser endomicroscopy (CLE), endoscopic ultrasound (EUS), photoacoustic endoscopy (PAE), and photoacoustic endoscope ultrasound miniprobe (PAE-UE) ~\cite{bhushan2019photoacoustic}. \\
\textbf{4) Ophthalmology Imaging Modalities:}
Ophthalmology is a branch of medical sciences that deals with the analysis, diagnosis, treatment and recovery of disorders associated with eyes. For the diagnosis and treatments of different eye parts, various types of imaging modalities are used which have different properties depending on what parts of eyes are considered and which types of diseases are targeted. The important imaging modalities used for the diagnosis and treatment of eyes include fundus photography, scanning laser ophthalmoscope (SLO), optical coherence tomography (OCT), and PAOM ~\cite{kumar2020complementary}.\\
5) \textbf{Dermatology Imaging:}
Dermatology is a branch of medical sciences that deals with the skin. The field has responsibility of both the medicinal purpose as well as surgical purpose. There are many imaging modalities that used for diagnosis and treatment of skin related diseases e.g., dermoscopy (related to skin), trichoscopy (related to scalp and hair), and onychoscopy (related to nails) ~\cite{hamblin2016introduction}.\\
\textbf{6) Functional Imaging:}
MRI, CT, US, positron emission tomography (PET), and single photon emission computed tomography (SPECT) are functional images because these are used for analysing brain functionalities ~\cite{elnakib2013developing}.\\
\textbf{7) Microscopic Imaging:}
The basic concept of microscopic imaging is that it uses microscope for capturing the images of different body parts at very deep and detailed cellular levels for the detection and diagnosis of various abnormalities in the organs. There are many types of microscopic imaging modalities but important of them include  selective plane illumination microscopy (SPIM), structured structure illumination microscopy (SSIM), two photon intravital microscopy (2P-IVM), two photon micro endoscopy (2P-ME) ~\cite{bullen2008microscopic}.\\
\textbf{8) Histopathology Imaging:}
Histopathology is the study of the signs of the disease using the microscopic examination of a biopsy or surgical specimen that is processed and fixed onto glass slides. The images used in this process are called histopathology images. To visualize different components of the tissue under a microscope, the sections are dyed with one or more stains. The histopathology images are used for the diagnosis of different human orgnas e.g., breast ~\cite{rong2023deep}.\\
\textbf{9) Molecular and Cellular Imaging}
Molecular and cellular images are very small types of images which are used for studying the finer details of structures of molecules and cells. These images are actually the microscopic images but used for the study of cells and molecules e.g., red blood celles (RBC), whie blood celles (WBC), yeast, crystal, and epithelium ~\cite{zhang2018panoptic}.

%% file: content/04-ObjDetections.tex
\section{DEEP LEARNING MODELS} 
\label{sec:DLModels}
The applications of deep learning in biomedical images and videos analysis have been the focus of research community due to their capability of processing complex data in an efficient manner ~\cite{cheng2021deep}. The biomedical image processing tasks by applying deep learning models are divided into four categories namely image classification, object detection, instance segmentation, and semantic segmentation. In semantic segmentation, each pixel inside the image is assigned to a specific class e.g., each pixel in liver will be assigned to tumour, parenchyma, or blood vessels. In case of instance segmentation, each individual object of a specific class is separately identified ~\cite{ren2015faster}. The object detection architectures are divided into three categories namely two-stage detection models ~\cite{redmon2016you}, single-stage detection models ~\cite{mittal2020deep}, and points-based detection models (which are also known as advanced detectors by many researchers). 
\begin{figure}[!htbp]
    \centering 
    \includegraphics[width=\linewidth]{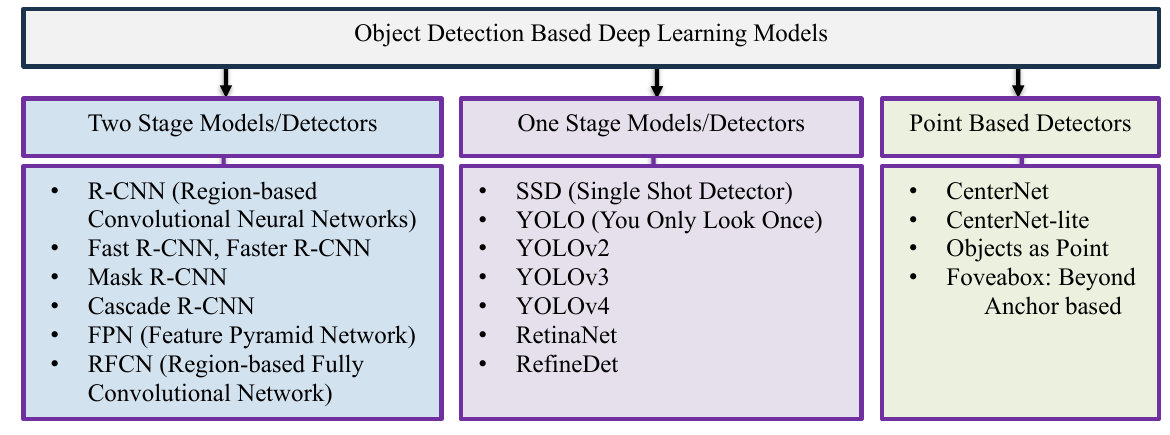}
    \caption{Object detection based deep learning models for biomedical imaging segmentation ~\cite{mittal2020deep}}
    \label{fig:DLf3}
\end{figure} 

The important two-stage detection approach has different models e.g., R-CNN, Fast-RCNN, Faster-RCNN, Mask-RCNN, Cascade-RCNN, FCN, and R-FCN. The one-stage well-known models are YOLO, SSD, YOLOv2/v3/v4, RefineDet, and RetinaNet. The points-based models include CornerNet, CornerNet-Lite, CenterNet, and FoveaBox as shown in~\Cref{fig:DLf3}. All are discussed in detail in the coming sections along with their structure, function, applications in biomedical imaging, and their pros and cons ~\cite{mittal2020deep}.
\subsection{Two-stage Detection Models}
\label{sec:2stge}
As explained in previous sections, the two-stage detection algorithms follow two steps for detection of specific object in the images in common, and boundary boxed areas for possible diseases in specific relevance to this work. In this section, the two stage detection models are discussed in detail. \\
\textit{1) Region Based Convolutional Neural Network (R-CNN):  } 
R-CNN is a two-stage object detection model used for the identification of different objects in an image. In case of biomedical imaging, it is used for the identification of different normal and diseased parts in the image. The R-CNN has been introduced to decrease the number of regions for reducing computation task. The detailed architecture and working mechanism of R-CNN are shown in ~\cite{girshick2014rich}.\\
\textit{2)	Faster R-CNN:}
Faster R-CNN is one of the most widely applicable two stage detection approach consisting of few important components. It has data layer, standard CNN, region proposal network (RPN) layer, Region of interest (ROI), ROI pooling layer, fully connected layer, and the classification layer. In the data layer, the input images are fed to the network for processing. The structure and function of faster R-CNN have been explained in ~\cite{xu2023improved}.
\textit{3)	Mask R-CNN:}
The mask R-CNN is a deep learning model used for object detection-based segmentation of images in computer vision. The output of mask R-CNN comes in three forms namely the class label of the object, the boundary box of the object, and the mask. The standard architecture of mask R-CNN consists of mainly three components which are the network backbone, region proposal network (RPN), and the ROIAlign. The tasks of each of these modules are outlined in ~\cite{podder2021efficient}. \\
\textit{4)	Cascade R-CNN:}
As highlighted by the authors in ~\cite{cai2018cascade}, cascade RCNN has been developed to resolve the issues of overfiting and intersection over union threshold value problem by training the detectors in a sequential manner with the increasing value of threshold in which the output of one detector is given as training to the next detector. This sequential training of detectors resolves both the issues of overfitting and the quality mismatch problem ~\cite{cai2018cascade}. \\
\textit{5)	Feature Pyramid Network (FPN):}
Detection of objects with varying scales is very difficult and challenging task especially for small objects. Pyramids of same image of different scales can be used for object detection but this is a very time-consuming task and requires a lot of memory. As an alternative, we can use pyramid of features of the image and this pyramid can be used for object detection. FPN has been created as a feature extractor for achieving high accuracy and speed of processing as explained in ~\cite{hui2019understanding}.   \\
\textit{6)	Region Based Fully Convolutional Network (R-FCN):}
R-FCN is a two-stage object detection model consisting of region-proposals and classification of regions. In R-FCN, the regions are extracted by region proposal network which fully connected convolutional neural network. In the working procedure of R-FCN, the image is read by the convolutional layer which create feature maps. The feature maps are used for creation of region proposal network which consists of many regions of interest. The convolutional layers after the feature maps creates each region of interest. After the extraction of region of interest, pooling is applied to generate the vote for each region of interest ~\cite{dai2016r}.\\
\subsection{One-stage Detection Models}
\label{sec:1stge}
In contrast to two-stage detection algorithms which consists of two stages namely region proposals and classification or object detection, the one stage detection algorithm carries out the detection task in only one stage. It uses the concept of anchors and grid box for the localization of the objects inside the image. There are many one-stage algorithms e.g., YOLO, SSD, RefineNet, and RetinaNet. All these models are discussed in this section with finer detail.\\
\textit{1)	You Only Look Once (YOLO):}
In the two-stage object detection algorithms e.g., Fast R-CNN, the model first identifies the regions of interest with the help of RPN and in the second stage each region of interest is given class labels for detection. In case of YOLO, both these functions are performed simultaneously with a fully connected convolutional layer. After YOLO was introduced in 2015, so far, few versions have been developed each having different pros and cons. The current variants of YOLO include YOLOV2 to YOLOV8. The working mechanism of YOLO consists of few stages~\cite{kavitha2023brain,chang2018automatic,xu2023automatic,zhong2019real,liu2023yolo,osama2023empowering,wu2023one}.\\ 
\textit{2)	Single Shot Detector (SSD):}
SSD is a one-stage object detection model consisting of two parts namely the backbone and the SSD head. The backbone is a type of pre-trained convolutional neural network used for feature extraction. The SSD head is one convolutional layer of few convolutional layers that generates the bounding box and class of the object. The complete working procedure of SSD is shown in ~\cite{xu2023dynamic,kulhare2018ultrasound,cao2017breast,souaidi2022new,chen2021single,zhang2018single}.  \\
\textit{3)	RefineDet:}
It is a single stage detection model that carries out the object detection process in only one stage. The simple structure of RefineDet consists of two parts namely anchor refinement module (ARM) and object detection module (ODM). There are pre-defined anchor boxes with fixed sizes, ratios and locations. The ARM tries to remove negative anchors for reducing search space to classify and adjust the sizes and locations of anchor boxes for better initialization for the regressor ~\cite{lin2017refinenet}. \\
\textit{4)	RetinaNet:}
It is another one-stage detection model consisting of one backbone and two subnetworks. The backbone is a convolutional neural network like ResNet used for feature map generation of the whole image. The first subnetwork performs the classification based on the outputs of backbone model whereas the second subnetwork is responsible for identification of bounding boxes and the class of each bounding box~\cite{ma2018novel,lin2017refinenet,ahmed2023new,rehman2022deep,lucena2020detection}.

\begin{figure*}[!htbp]
    \centering 
    \includegraphics[width=1\linewidth]{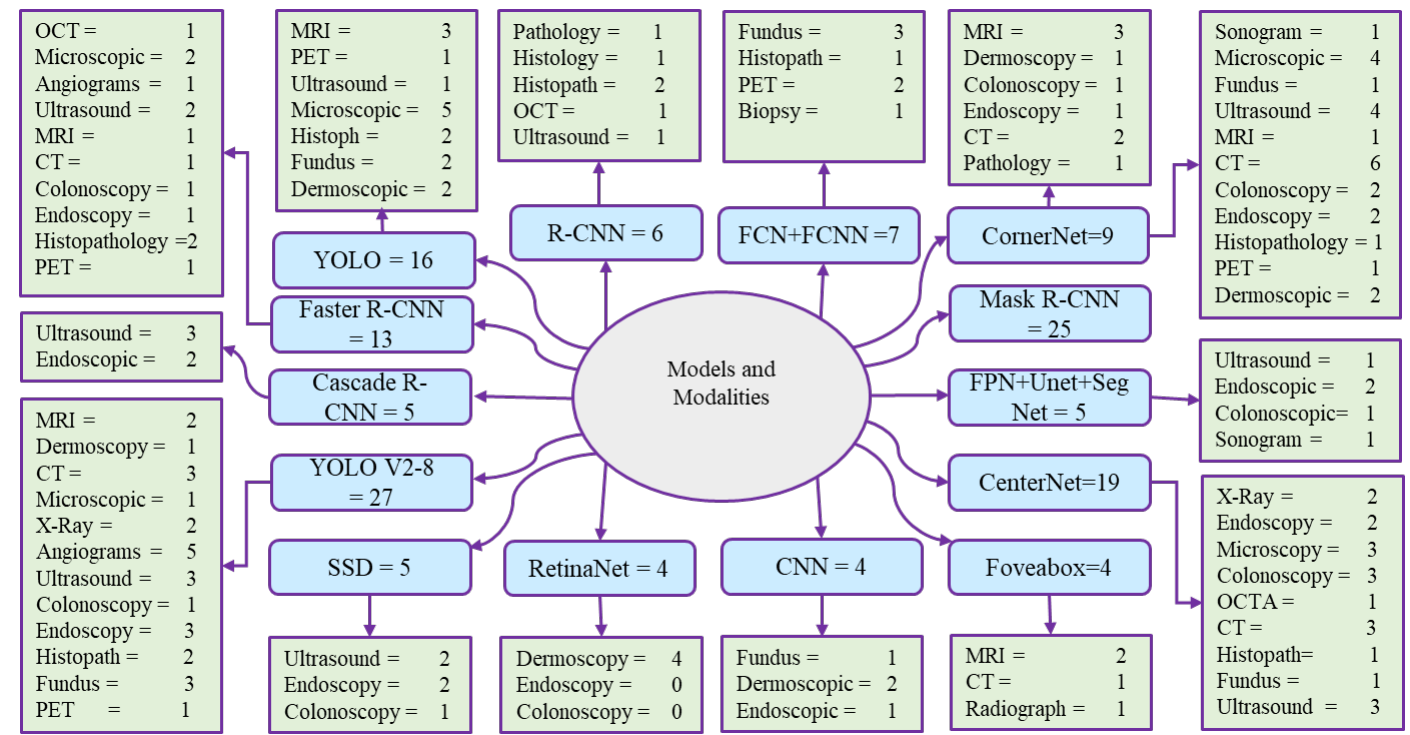}
    \caption{A chart showing the models (light blue color) and modalities (light green color). The numeric values indicates the number of papers who have used these models/modalities. }
    \label{fig:ModelsANDModalities}
\end{figure*} 
\subsection{Point Based Detection Models}
\label{sec:point}
There are some object detection algorithms which can applied in procedures sightly different both from one-stage detection techniques as well as two-stage detection techniques. All these algorithms have been placed in a separate category. Their working mechanisms and power are higher than both the already discussed two-stage detectors as well as one-stage detectors. Therefore, they have been categorized as point based detectors because they use specific points for detection of objects in images. Some of these detectors are discussed in this section\\
1) CornerNet:
In cornerNet, the bounding boxes of objects inside image are detected using pair key points namely top-left corner and bottom right corner. This process of finding keypoints is performed by using convolutional neural network. By using this procedure, the need for designing anchor boxes which are applied in previously discussed one-stage detecting approaches is eliminated ~\cite{law2018cornernet}.\\
2)	CornerNet-Lite:
CornerNet-Lite is a combined architecture of two different variants of standard cornerNet namely CornerNet-Saccade and CornerNet-Squeeze ~\cite{law2019cornernetlite}. The first variant applies the concept of attention mechanism for eliminating the need of processing all pixels of an image whereas the main responsibility of the CornerNet-Squeeze is to introduce a powerful backbone architecture.\\ 
3)	CenterNet:
In this architecture, the objects are identified by finding their central points. From these central points, the other properties of objects like orientation, location, size, and even pose of objects are found. The resulting architecture also known as CenterNet is faster, simpler, and more accurate than the traditional bounding box-based object detection models ~\cite{zhou2019objects}.\\ 
4) Foveabox:
FoveaBox is flexible, accurate, and anchor free architecture for object detection. All the previous anchor-based object detection models utilize anchors to detect objects inside an image, their performance and ability of generalization are limited to the design of anchors. In contrast, the Foveabox directly identifies boundary boxes and the existence of objects in the image without refereeing to the pre-defined anchors ~\cite{kong2020foveabox}.

%% file: content/05-Results.tex
\section{SELECTED ARTICLES AND ANALYSIS}
\label{sec:mainSegementation}
\begin{figure*}[tbp]
    \centering
    \includegraphics[width=\linewidth]{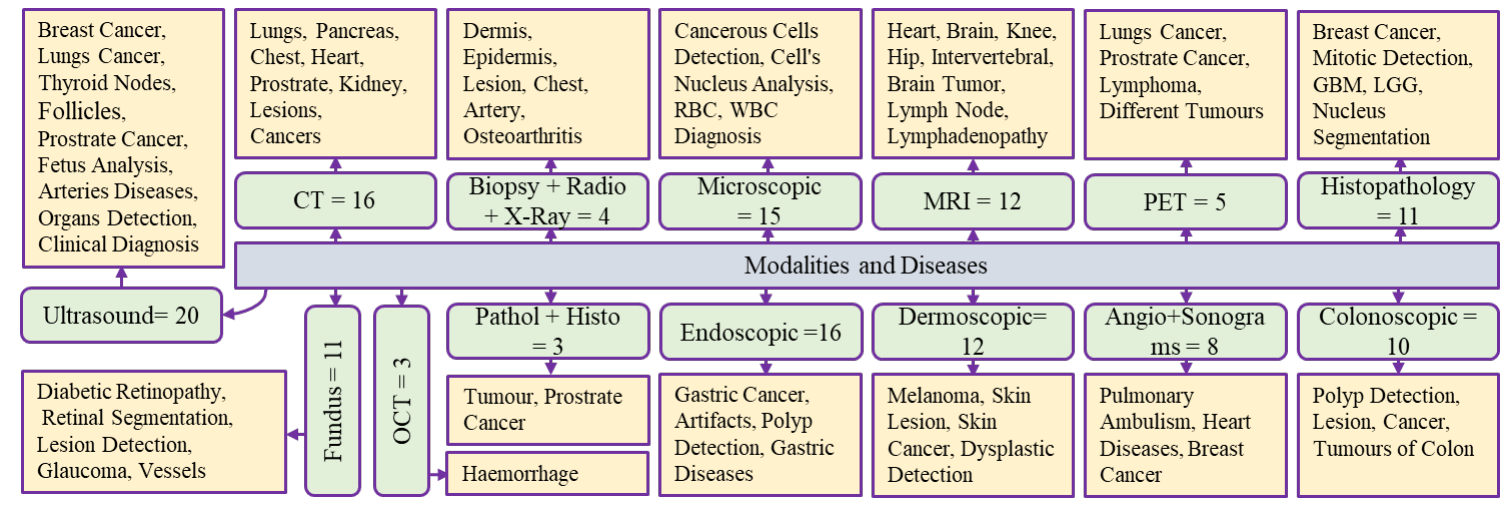}
    \caption{A chart showing the modalities (light green) and diseases (yellow). The numeric values indicates the number of papers with each modality. }
    \label{fig:ModalitiesANDdiseases}
\end{figure*}

\begin{figure}[!htbp]
    \centering
    \includegraphics[width=\linewidth]{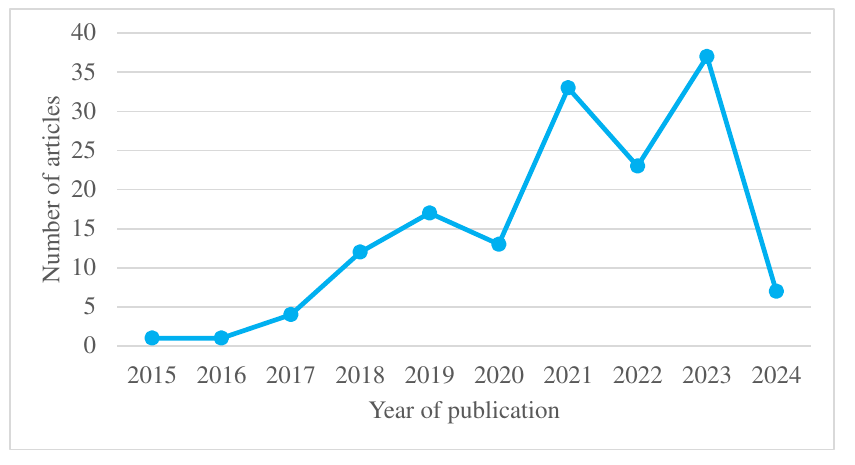}
    \caption{Articles distribution on the basis of year of publication. }
    \label{fig:yearwisedistribution}
\end{figure}

 In this section, we discuss all the selected articles with in-depth analysis of the extracted information from each article. Each article is briefly described with its pros and cons. The articles are classified in three categories including two-stage detection, one-stage detection, and point based. In addition, modalities and diseases concerned with each model are discussed in detail. The challenges faced by these models, and their pros and cons have also been presented. \Cref{fig:ModelsANDModalities} shows different models and concerned modalities in a chart; where as \Cref{fig:ModalitiesANDdiseases} present a chart of the modalities and concerned diseases. \Cref{fig:yearwisedistribution} shows the year-wise distribution of the selected articles.
\subsection{Articles related to Two-stage Detection}
As described in~\cref{sec:2stge}, two-stage object detection models include R-CNN, Fast R-CNN, Faster R-CNN, Mask R-CNN, Cascade R-CNN, FPN, and R-FCN. This section discusses the application of these models for the segmentation of biomedical images.\\
For example, a modified CNN called  region-based convolutional neural network (R-CNN) multi-task model is used for the segmentation of histological images to detect prostate cancer~\cite{li2018path}, and for the detection of subretinal hemorrhage in OCT images ~\cite{suchetha2021region}. MITOS-RCNN~\cite{rao2018mitos} is another approach  that uses R-CNN for the detection of breast cancer in the histopathological images. The R-CNN has also been used for segmenting pathological images to detect various tumors such as Glioblastoma Multiforme (GBM), Head and Neck Squamous Cell Carcinoma (HNSCC), Non-Small Cell Lung Cancer (NSCLC), and Lower Grade Glioma (LGG)~\cite{ma2022tensor}.  Similarly, mask scoring R-CNN~\cite{lei2021breast} is used for breast cancer detection in automatic breast ultrasound (ABUS) imaging. Zhang et al. ~\cite{ma2021real} proposed an end to end cell R-CNN for the analysis of pathological images. Their segmentation is based on global convolutional neural network (GCN).

  Two-stage CNN model~\cite{van2024segmentation} is used for the segmentation of hard exudate lesions in colour fundus images. Similarly, a combination of convolutional neural network and the level sets was proposed by Huang et al ~\cite{huang2024skin} is used for the segmentation of skin images for the detection of lesions. Yuan and Lo ~\cite{yuan2017improving} proposed an enhanced version of convolutional neural network for the segmentation of dermoscopic images to detect lesions in the skin. The faster R-CNN has been used in another article~\cite{yang2017faster} for the detection of cells in the microscopic images. Another alternative enhanced model of faster R-CNN~\cite{yuan2021improved} has been used for segmentation of angiography images to detect pulmonary embolism. Similarly, Huang-Nan et al ~\cite{huang2022image} used fast R-CNN in combination with transfer learning  for foot images to detect diabetic foot ulcers. Faster R-CNN has also been used for the detection of polyp in the colonoscopy images~\cite{chen2021self},  the segmentation of ultrasound images to detect thyroid nodules~\cite{tianlei2023improved}, segmentation of breast cancer~\cite{harrison2021tumor},  , and segmentation of oncologic FDG PET images to detect lesion ~\cite{gao2019automatic}. \\
  Recently, a hybrid model~\cite{cui2023improved} composed of 3D U-Net and Faster R-CNN is proposed which has been used for the segmentation of heart CT and MRI images for the detection of cardiovascular diseases.  Krenzer et al. ~\cite{krenzer2020endoscopic} applied three different deep learning models for the detection and segmentation of gastroenterological diseases in the endoscopy images.  Khan et al. ~\cite{khan2023smdetector} proposed a model named SMDetector which is based on faster R-CNN for the mitotic detection in the histopathological images of breast cancer. Similarly, Zhang et al. ~\cite{zhang2016cancer} suggested the use of faster R-CNN for the segmentation of microscopic images to detect cancer cells. The authors have applied faster R-CNN and the circle scanning algorithm for improving the detection of cancer cells. An enhanced mask R-CNN was applied by the authors in work presented in ~\cite{shu2020improved}. The proposed model was applied to 4341 CT images of heart, lungs, CTV, and PTV with data division of 80 $\%$ for training and 20 $\%$ for testing. In another study, mask R-CNN was used for the image segmentation to find different oral diseases. The model was applied to two sets of oral images for detection of two conditions being cold sore and canker sore ~\cite{anantharaman2018utilizing}. Similarly, the mask R-CNN was applied for the detection of gastric cancer in the endoscopic images ~\cite{shibata2020automated}. In another attempt, the authors used improved version of mask R-CNN for the segmentation of MRI images for knee and hip assessments ~\cite{felfeliyan2023self}. \\
Liu and Li~\cite{liu2018mask} applied mask R-CNN for the segmentation of ultrasound images. They introduced some modification in the RPN for improving the segmentation accuracy of the mask R-CNN model. Similarly, a combination of mask R-CNN and 3D U-Net is used for the segmentation of CT images of pancreas~\cite{dogan2021two}. Furthermore, mask R-CNN has also been used for segmentation of lungs nodule~\cite{liu2018segmentation}, and 3D visualization of pulmonary nodule ~\cite{cai2020mask}.  The proposed model~\cite{liu2018segmentation} consists of backbone network, RPN, ROI Align and fully convolutional neural network. The authors in ~\cite{mulay2020liver} proposed a combination of mask R-CNN and holistically nested edge detection (HED) model for the segmentation of MRI and CT images of liver.\\
 Recently, Zhiyong et al. ~\cite{liu2021deep} proposed a hybrid model of mask R-CNN and inception-V3 for the segmentation of ultrasound images to detect prostate cancer. In the proposed model sobel convolution has been applied for sharpening the input images. Similarly, Yuanhao et al.~\cite{liang2019simultaneous} performed the classification and segmentation of ultrasound images for detection of breast tumour. They applied superpixel elastic deformation approach for data augmentation and mask R-CNN for segmentation of the ultrasound images for possible identification of breast tumour. Another study used mask R-CNN architecture for the segmentation of sonogram images to detect breast cancer~\cite{chiao2019detection}.  This architecture~\cite{chiao2019detection} includes a CNN for feature map generation, an RPN for extracting regions of interest, and ROI Align for extracting features of each ROI.  In the proposed model, the RPN has been combined with fast R-CNN.\\
 For assessing fetal growth in ultrasound images, an advanced version of mask R-CNN, known as Mask R2-CNN~\cite{moccia2021mask} is used. Furthermore, mask R-CNN has been employed for the segmentation and detection of adenomatous colorectal polyps in colonoscopy images~\cite{meng2020automatic}, the detection of early gastric cancer in endoscopy images~\cite{jin2022automatic}, the automatic segmentation of nuclei in histopathology images~\cite{jung2019automatic}, and the segmentation of fundus images to detect lesions in diabetic retinopathy~\cite{santos2023new}. 
\begin{table*}[htbp]
 \label{tabVI:medical_image_analysis_studies}
  \centering
  \caption{Two-Stage Detectors for Biomedical Image Segmentation.}
\resizebox{1.05\linewidth}{!}{
 \begin{tabular}{|c|c|p{3.25cm}|p{3.95cm}|p{5.4cm}|p{6.0cm}|p{5cm}|}
    \hline
    \textbf{ID-Ref} & \textbf{Year} & \textbf{Modality} & \textbf{Model} & \textbf{Disease $|$ Organ} & \textbf{Evaluation Metrics} & \textbf{Results} \\
    \hline
    A001~ ~\cite{ma2022tensor} & 2022 & Pathological & R-CNN & Tumours & F1-Score, Dice, Hausdorff & 0.8216, 0.7088, 11.3141 \\
    \hline
     A002~~\cite{li2018path} & 2018 & Histology & R-CNN & Prostate Cancer & IoU, Accuracy & 79.56\%, 89.40\%\\
    \hline
     A003~~\cite{suchetha2021region} & 2021 & OCT & R-CNN, Faster R-CNN & Subretinal Haemorrhage & Sensitivity, Specificity, Accuracy & 90.36\%, 85.98\%, 93.98\% \\
    \hline
    A004~~\cite{lei2021breast} & 2021 & Ultrasound & R-CNN & Breast Cancer & DSC, JI, Hausdorff, MSD, RMSD, CMD & 85.0\%, 75.2\%, 1.65, 0.49, 0.76, 0.67 \\
    \hline
    A005~~\cite{ma2021real} & 2021 & Histopathological Images & R-CNN & GBM, HNSCC, LGG, NSCLC & F1-Score, Dice, Hausdorff & 0.8216, 0.7088, 11.3141 \\
    \hline
    A006~~\cite{rao2018mitos} & 2018 & Histopathology & R-CNN & Breast Cancer & Recall & 0.955 \\
    \hline
    A007~~\cite{van2024segmentation} & 2024 & Fundus & CNN & Exudate Lesions & AUPRC, DSC, IOU & 0.893, 76.6\%, 62.4\% \\
    \hline
    A008~~\cite{huang2024skin} & 2024 & Dermoscopic Images & CNN & Skin Lesions & Jaccard Index & 77.0\%, 85.1\% \\
    \hline
    A009~~\cite{yuan2017improving} & 2017 & Dermoscopic Images & CNN & Skin disease & Jaccard Index & 0.765 \\
    \hline
    A010~~\cite{yang2017faster} & 2017 & Microscopic Images & Faster R-CNN & RBC, WBC, Yeast, etc. & Recall, Precision, F1-Score & 0.9773, 0.7768, 0.8656 \\
    \hline
    A011~~\cite{yuan2021improved} & 2021 & Angiograms & Faster R-CNN & Pulmonary Embolism & AP, Sensitivity, Specificity & 85.88\%, 87.30\%, 86.29\% \\
    \hline
    A012~~\cite{huang2022image} & 2022 & JPG Format + CSV files & Fast R-CNN & Diabetic Foot Ulcers & Accuracy & 90\% \\
    \hline
    A013~~\cite{tianlei2023improved} & 2023 & Ultrasound & Faster R-NN & Thyroid Nodules & AP 50\%, AP 75\%, AP 50:5:95\% & 97.4\%, 81.3\%, 68.7\% \\
    \hline
    A014~~\cite{cui2023improved} & 2023 & CT MRI & Faster R-CNN U-Net & Heart Disease & Dice, Jaccard, Hausdorff, Running Time & 0.864, 0.767, 29.918, 4.1 Sec \\
    \hline
    A015~~\cite{chen2021self} & 2021 & Colonoscopy & Faster R-CNN & Polyp detection & Precision, Recall, F1-Score & 0.943, 0.925, 0.934 \\
    \hline
    A016~~\cite{krenzer2020endoscopic} & 2020 & Endoscopy & Faster R-CNN Cascade R-CNN YOLOv3 & Gastric Disease & mAP & 37.20, 52.44, 44.49 \\
    \hline
    A017~~\cite{harrison2021tumor} & 2021 & Histopathology & Faster R-CNN & Breast Cancer & Precision, Sensitivity, IoU, F1 Score & 1, 0.03, 0.55, 0.05 \\
    \hline
    A018~~\cite{gao2019automatic} & 2019 & PET & Faster R-CNN & Lesion Detection & Accuracy, Sensitivity, Specificity, PPV, NPV & 97.2\%, 95.4\%, 98.1\%, 96.8\%, 97.0\% \\
    \hline
    A019~~\cite{khan2023smdetector} & 2023 & Histopathology & Faster R-CNN & Mitotic Detection & AP, AP 50\%, AP 75\%, AP small, AP medium, AP large & 25.35, 50.31, 25.45, 17.6, 16.17, -1 \\
    \hline
    A020~~\cite{zhang2016cancer} & 2016 & Microscopic Images & Faster R-CNN & Cancer Cells & Precision, Recall, AP, AUC & 0.990, 0.990, 0.906, 0.661 \\
    \hline
    A021~~\cite{shu2020improved} & 2020 & CT & Mask R-CNN & Heart, Lungs & DICE, JI & 97.6\%, 98.1\%, 95.1\%, 97.8\% \\
    \hline
    A022~~\cite{anantharaman2018utilizing} & 2018 & Oral Images & Mask R-CNN & Cold Sore, Canker, Sore & Dice & 0.774, 0.714 \\
    \hline
    A023~~\cite{shibata2020automated} & 2020 & Endoscopic Images & Mask R-CNN & Gastric Cancer & Image Sensitivity Lesion Sensitivity & 0.76, 0.96 \\
    \hline
    A024~~\cite{felfeliyan2023self} & 2023 & MRI & Mask R-CNN & Knee, Hip & Dice & 0.76, 0.75 \\
    \hline
    A025~~\cite{liu2018mask} & 2018 & Ultrasound & Mask R-CNN & Follicles & Accuracy & 96.5\% \\
    \hline
    A026~~\cite{dogan2021two} & 2021 & CT & U-Net, Mask R-CNN & Pancreas & DSC, JI, REC, ACC & 86.15\%, 75.93\%, 86.27\%, 99.95\% \\
    \hline
    A027~~\cite{liu2018segmentation} & 2018 & CT & Mask R-CNN & Lungs & mAP & 0.7334 \\
    \hline
    A028~~\cite{mulay2020liver} & 2019 & MRI CT & Mask R-CNN & Liver & Dice & CT: 0.94 MRI: 0.91 \\
    \hline
    A029~~\cite{cai2020mask} & 2020 & CT & Mask R-CNN & Chest & Sensitivity, FP & 88.7\%, 88.1\% \\
    \hline
    A030~~\cite{liu2021deep} & 2021 & Ultrasound & Inception V3 + Mask R-CNN & Prostate Cancer & mAP, DICE, IOU, AP & 0.88, 0.87, 0.79, 0.92 \\
    \hline
    A031~~\cite{liang2019simultaneous} & 2019 & Ultrasound & Mask R-CNN & Breast Cancer & DICE, Accuracy, TP, TN & 83.93\%, 80.42\%, 63.64\%, 87.88\% \\
    \hline
    A032~~\cite{chiao2019detection} & 2019 & Sonogram & Mask R-CNN & Breast Cancer & mAP & 0.75 \\
    \hline
    A033~~\cite{moccia2021mask} & 2021 & Ultrasound & Mask R-CNN & Fetal growth & AD, DI, DSC, H & 1.95, -0.31, 97.90\%, 1.5 \\
    \hline
    A034~~\cite{meng2020automatic} & 2020 & Colonoscopy & Mask R-CNN & Adenomatous colorectal polyps & mAP 50\%, mAP 70\%, mAP 75\% & 89.50\%, 78.40\%, 73.50\% \\
    \hline
    A035~~\cite{jin2022automatic} & 2022 & Endoscopy & Mask R-CNN & Gastric Cancer & Accuracy, Sensitivity, Specificity, PPV, NPV & 90.25\%, 91.06\%, 89.01\%, 92.61\%, 86.81\% \\
    \hline
    A036~~\cite{qadir2019polyp} & 2019 & Colonoscopy & Mask R-CNN & Polyp & Precision, Recall, Jaccard, Dice & 94.1\%, 86.1\%, 73.2, 80.19 \\
    \hline
    A037~~\cite{jung2019automatic} & 2019 & Histopathology & Mask R-CNN & Nuclei Segmentation & F1 Score, Precision, Recall, Dice & 0.913, 0.907, 0.923, 0.835 \\
    \hline
    A038~~\cite{santos2023new} & 2023 & Fundus & Mask R-CNN & 4 Different Lesions & AP, mAP & 0.2515, 0.1548, 0.1042, 0.1577, 0.2687, 0.1589, 0.1274, 0.1388 \\
    \hline
    A039~~\cite{huang2020skin} & 2020 & Dermoscopic Images & Mask R-CNN & Skin Disease & Precision, Recall, Average Precision & 0.9021, 0.9187, 0.9085 \\
    \hline
    A040~~\cite{bagheri2021skin} & 2021 & Dermoscopic Images & Mask R-CNN RetinaDeepLab & Skin Disease & Jaccard & 80.04\% \\
    \hline
    A041~~\cite{zhang2019multiscale} & 2019 & PET & Mask R-CNN & Lungs Tumour & Precision, Recall, F1-Score & 0.90, 1, 0.95 \\
    \hline
     A042~~\cite{wang2021prostate} & 2021 & PET CT & Mask R-CNN & Tumour Prostrate & ACD, VD, DSC & 0.83+-0.91, -0.01+-+-0.79, 0.84+-0.09 \\
    \hline
    A043~~\cite{wang2024deep} & 2024 & Microscopic Images & Mask R-CNN & Cell Detection & Dice, Accuracy, Precision, Sensitivity, F1 Score, Specificity, Volume Difference & 0.862, 0.945, 0.901, 0.827, 0.862, 0.977, 0.082 \\
    \hline
    A044~~\cite{hoorali2023automatic} & 2023 & Microscopic Images & Mask R-CNN & Anthrax Tissue Disease & AP, AP75, IoU, DICE & 0.560, 0.613, 0.751, 0.840 \\
    \hline
    A045~~\cite{chen2021celltrack} & 2021 & Microscopic Images & Mask R-CNN & Cell Segmentation & SEG, DET, TRA, D-T & 85.26\%, 97.14\%, 97.05\%, 0.09 \\
    \hline
    A046~~\cite{wang2023deep} & 2023 & Ultrasound & Cascade R-CNN, Faster R-CNN & Arteries Disease & Accuracy, AO, NRA, RASIRA & 86.5\%, 90.4\%, 84.7, 88.8\% \\
    \hline
    A047~~\cite{zheng2022automated} & 2022 & Ultrasound & Cascade R-CNN & Thyroid Nodules & mAP, TPR, FPR, TNR, FNR, Accuracy & 87.1\%, 0.9841, 0.0082, 0.9894, 0.0158, 0.9867 \\
    \hline
    A048~~\cite{chen2022cnet} & 2022 & Ultrasound & Cascade R-CNN & Breast Cancer & Accuracy, Jaccard, Precision, Recall, Specificity & 97.51\%, 72.47\%, 82.70\%, 99.13\%, 79.35\% \\
    \hline 
    A049~~\cite{zhang2019detection} & 2019 & Endoscopy & Cascade R-CNN FPN & Multi-class Artifacts & Score Detection, Score Segmentation & 0.3429, 0.3500 \\
    \hline
    A050~~\cite{yang2019endoscopic} & 2019 & Endoscopy & Cascade R-CNN & Artifacts & mAP, IoU, Score & 0.3235, 0.4172, 0.3610 \\
    \hline
    A051~~\cite{chen2023ffpn} & 2023 & Ultrasound & FPN & Fetal head segmentation Heart & DICE, IoU, Hausdorff, Conformity & 89.08\%, 80.70\%, 19.76, 74.64\% \\
    \hline
    A052~~\cite{su2023fednet} & 2023 & Endoscopy & FPN & Polyp Different Varying &different & varying \\
    \hline
    A053~~\cite{wuyang2021joint} & 2021 & Endoscopy & FPN + CNN & Polyp & Average Precision, Segmentation Score & 0.8986, 0.7771 \\
    \hline
    A054~~\cite{su2015region} & 2015 & Histopathology & FCNN & Breast Regions & Average Precision, Average Recall, Average F1 Score, Variance Precision, Variance Recall, Variance F1 Score & 0.91, 0.82, 0.85, 0.015, 0.02, 0.01 \\
    \hline
    A055~~\cite{xue2019deep} & 2019 & Fundus & FCNN & Diabetic Retinopathy & DSC, Sensitivity, Specificity, Accuracy & 96.7\%, 98.1\%, 99.9\%, 99.9\% \\
    \hline
    A056~~\cite{li2018vessel} & 2018 & Fundus & FCNN & Vessels Detection & Accuracy, Sensitivity, Specificity & 0.9210, 0.7215, 0.9576 \\
    \hline
    A057~~\cite{feng2017deep} & 2017 & Fundus & FCNN & Retinal Segmentation & Sensitivity, Specificity, PPV, F1 Score & 93.12\%, 99.56\%, 89.90\%, 90.93\% \\
    \hline
    A058~~\cite{pal2018psoriasis} & 2018 & Biopsy & FCN & Dermis Epidermis & RCPC, Jaccard & 0.8801, 0.8845, 0.8900, 0.9962 \\
    \hline
    A059~~\cite{zhao2018tumor} & 2018 & PET CT & FCN & Tumour & DSC, Classification Error, Volume Error & 0.85, 0.33, 0.15 \\
    \hline
    A060~~\cite{pang2024mtrpet} & 2024 & PET & FCN & Lymphoma & Detection Sensibility, Dice & 0.9953, 0.8665 \\
    \hline
    A061~~\cite{jha2020doubleu} & 2020 & Dermoscopy Colonoscopy Endoscopy & U-Net & Multiple & mIoU, DSC, Recall, Precision & 0.6255, 0.7649, 0.7156, 0.8007 \\
    \hline
    A062~~\cite{yang2023rapid} & 2023 & Sonograms & D-CNN U-Net & Breast Cancer & Accuracy, Sensitivity, Specificity, AUC, UIoU, MDice & 97\%, 97.7\%, 96.4\%, 0.96, 0.89, 0.92 \\
    \hline
    A063~~\cite{gu2019cenet} & 2019 & Angiographic CT MRI & CE-Net & Lungs, Vessels, Eyes & Sensitivity, Accuracy, AUC & 0.8309, 0.9545, 0.9779 \\
    \hline
    A064~~\cite{jha2021realtime} & 2021 & Colonoscopy & ColonSegNet & Polyp & AP, IoU, DSC, Jaccard, Precision, Recall & 0.8000, 0.8100, 0.7239, 0.8435, 0.8496, 0.9493 \\
    \hline
  \end{tabular}
  }
\end{table*}
It has been used for the segmentation of dermatology images to detect lesions in the skin~\cite{huang2020skin,bagheri2021skin}, for the segmentation of PET images to detection tumour in human lungs~\cite{zhang2019multiscale}, and for the analysis of microscopic images~\cite{wang2024deep}. Few studies improved mask R-CNN with some modifications. For example, a dual mask R-CNN is used for the segmentation of PET and CT images to detect tumour and prostrate~\cite{wang2021prostate}. 
 Another enhanced version of mask R-CNN called URCNN~\cite{hoorali2023automatic} is used for the segmentation of microscopic images to detect anthrax and some other tissue diseases. Similarly, another variant is a faster R-CNN~\cite{chen2021celltrack} which is used for the segmentation of microscopic images to detect cancer cells.  Further R-CNN's improved variants include  cascade R-CNN~\cite{chen2022cnet} used for the segmentation of ultrasound images to detect thyroid nodules, another variant is used for the segmentation of breast ultrasound images to detect lesions~\cite{zheng2022automated}, a hybrid model of R-CNN and pyramid network (FPN) for the detection and segmentation of different multi-class artifacts in the endoscopy images~\cite{zhang2019detection}, and a cascade R-CNN is used for the detection and segmentation of artefacts in the endoscopic images~\cite{yang2019endoscopic}. \\
 FPN has also been improve to improve varients including Fourier feature pyramid network (FFPN) used for for the segmentation of ultrasound images~\cite{chen2023ffpn},   FedNet for the segmentation of polyp in the endoscopic images~\cite{su2023fednet}, and FPN with other CNN networks for the detection and segmentation of polyp in the endoscopy images ~\cite{wuyang2021joint}. Su et al. ~\cite{su2015region}, suggested an improved version of CNN named fCNN standing for fast scanning deep convolutional neural network for the segmentation and identification of regions in histopathology images of breast.
A recent article~\cite{wang2023deep} has tested three two-stages model namely cascade R-CNN, faster R-CNN, and double head R-CNN on ultrasound images of arteries for their segmentation to detect various abnormalities. Another study a multi-task segmentation approach for the segmentation of retina for the detection of diabetic retinopathy~\cite{xue2019deep}. Similarly, the fully connected convolutional neural networks have been used for the segmentation of retinal images~\cite{li2018vessel,feng2017deep},  for the segmentation of PET images ~\cite{pang2024mtrpet,zhao2018tumor} and CT images to detect tumour~\cite{zhao2018tumor}. Another approach is the deep convolutional neural network  for the segmentation of biopsy images to detect dermis, epidermis, and non-tissue regions~\cite{pal2018psoriasis}.  \\
Double U-Net~\cite{jha2020doubleu} is an improved version of the standard U-Net  in which two U-Nets are stacked on the top of each other. The model was evaluated with four publicly available dataset of different modalities e.g., dermoscopy, colonoscopy, and microscopy~\cite{jha2020doubleu}. Attention U-Net and deep convolutional neural network have been used for the segmentation and classification of sonography ultrasound images for the detection of breast cancer~\cite{yang2023rapid}. CE-Net (context encoder network)~\cite{gu2019cenet} is another model which is applied for the segmentation of 2D medical imaging. Similarly, ColonSegnet~\cite{jha2021realtime} is used for the detection, localization, and segmentation of polyp in the colonoscopy images.
\subsubsection{Analysis of Two Stage Detectors}
Different imaging modalities that have been the target of two stage detectors include CT images, MRI images, X-rays, ultrasound images, endoscopic images, colonoscopic images, dermoscopic images, histopathology images, pathological images, microscopic images, fundus images, and PET images.  In the same fashion, the most frequently applied two stage detectors include mask R-CNN, Faster R-CNN, FCNN, FPN, cascade R-CNN, and RetinaNet, respectively. The most powerful, simple, accurate, and easily implementable models of these architectures is the mask R-CNN. If we critically analyse, the mask R-CNN model provide better results than other two stage detection models. In some cases, the faster R-CNN also provides comparatively better results than other models. Similarly, some fluctuations can also be observed in the results of other two stage detection models. The major factors behind these variations in the results of all models include the size of datasets been processed, the imaging modality, the architecture of the detection model, the pre-processing and post processing mechanisms applied, and the hardware and software specifications used in the experimentation.

\subsection{Articles Related to One-Stage Detection}
One-stage object detection models used for the segmentation of biomedical images include YOLO, SSD, YOLOv1/v2/v3/v4, RefineNet, and RetinaNet. In this section, these models, as applied to biomedical image segmentation, are discussed in detail along with other relevant considerations. In this regard,   
YOLO~\cite{afshari2018automatic} has been used to localize and detect various active organs in 3D PET imaging modalities. YOLOv3 has been applied for the segmentation of MRI images to detect herniated intervertebral discs~\cite{zhong2019real}, and for the segmentation of 2D and 3D CT images to localize kidneys and identify conditions such as cystic kidneys, hypertrophied kidneys, and tumoral kidneys~\cite{lemay2019kidney}. Other studies have utilized YOLO~\cite{kavitha2023brain} and YOLOv5~\cite{suryawanshi2024brain} for the segmentation of brain MRI images to detect tumors.  
Similarly, YOLOv4 has been used for the segmentation of ultrasound images to detect various organs~\cite{xie2022endoscopic}, YOLOv7 for the segmentation of echocardiographic images to identify and segment the anatomical structures of the left heart~\cite{mortada2023segmentation}, and YOLOv8 for processing X-ray angiography to segment the coronary artery~\cite{liu2023yolo}. Furthermore, the scaled-YOLOv4 has been applied for the detection of pulmonary embolism in CT angiograms~\cite{xu2023automatic}, and YOLOv5~\cite{wu2023one} has been used for multi-type lesion detection in angiography images. Santos et al. ~\cite{santos2022new} used YOLOV5 for lesion detection in fundus images.
In another study, YOLOv8 model is used for the segmentation of angiography images to identify and detect coronary artery stenosis~\cite{osama2023empowering}. Pkhrel et al. ~\cite{pokhrel2023data} used YOLOv8 along with the augmentation for the segmentation of coronary artery in the angiograms to detect diseases. \\
Some researchers have combined YOLO with other models to enhance performance. For instance, Chang et al.\cite{chang2018automatic} employed YOLO alongside a fully convolutional neural network for the segmentation and classification of heart MRI images. YOLO, in combination with the GrabCut model, has been used for the segmentation of dermoscopic images to detect skin lesions\cite{unver2019skin}. Moreno et al.~\cite{moreno2019combined} applied a combination of two convolutional neural networks for the segmentation of the left ventricle in heart MRI images. Other hybrid models include the YOLOv2 and Fourier Photography Microscopy (FPM) in combination for the detection of WBC ~\cite{wang2018soyolo}. 
The Faster R-CNN and YOLO are combined and used for the segmentation of chest X-Ray images to detect different abnormalities~\cite{nguyen2023anomalies}. YOLO-RF~\cite{lalitha2021detection} is a hybrid model of YOLO and random forest which has been used for diabetic detection. \\
YOLOv3-arch~\cite{pang2019novel} is another improvement of the YOLOv3  which has been used for the segmentation of CT images to identify cholelithiasis and classify gallstones.    Zhuang et al. ~\cite{zhuang2021automatic} used YOLOv3 for the segmentation of left ventricle in the echocardiography. Similarly, Zhang et al. ~\cite{zhang2021automated} suggested the use of and enhanced model YOLOV3 one-stage segmentation model for segmentation of ultrasound images to detect the location of thyroid nodules. Similarly, YOLO-CSE~\cite{hussein2023myositis} is an improvement of YOLOV5 which is used for the segmentation of muscle ultrasound to detect myositis. YOLOX~\cite{lu2022yolox} is used for the segmentation of ultrasound images to examine the fetus for congenital heart disease. Similarly, Ghose et al. ~\cite{ghose2023improved} suggested the use of an enhanced version of YOLOV5 to segment the colonoscopic images for the detection of polyp. YOLOV4-tiny~\cite{doniyorjon2022improved} has been used for the segmentation of endoscopic images to detect polyp.
Further studies  applied YOLOV4 for the segmentation of endoscopic images to detect multi-class artefact~\cite{kirthika2021yolov4}, and YOLOV2 for the detection of breast cancer in histopathological breast images~\cite{alzorgani2022deep}. YOLO has been for the segmentation of nucleus and characterization of tumour microenvironment~\cite{rong2023deep}, and for detection of nucleus and U-Net for the segmentation of nucleus in fluorescence microscopy~\cite{narotamo2019segmentation}.  YOLO has also been used for the identification of region proposal networks to diagnose cancer in breast histopathology images~\cite{bal2021yolo}. \\
Durak et al. ~\cite{durak2021deep} implemented different types of deep learning models for the detection of polyp in the gastro endoscopic images. Other different deep learning models including mask R-CNN, U-Net, CNN, and YOLO have been used for the segmentation of nucleus in the histopathological images~\cite{ayaz2021segmentation}. Similarly, CNN combined with YOLO has been used for the detection of glaucoma in the color fundus images~\cite{saha2023fast}.
Nair et al. ~\cite{nair2021mitotic} applied YOLOV4 for the nucleus detection in the histopathology images of breast. Similarly, Pal et al. ~\cite{pal2020detection} applied YOLOV3 for detection of lesions in the retinal fundus images.
 \\
   Other articles demonstrate the use of different models for more diseases. For example, YOLOv3 has been used for the localization of the optic disc in retinopathy fundus images~\cite{ramachandran2020fully}, YOLO for lesion detection in skin using dermoscopic images~\cite{bagheri2020two}, and fuzzy logic combined with object detection models for cancer detection in dermoscopic images~\cite{singh2023fuzzy}. Additionally, the YOLOv3 network has been applied for the segmentation of PET/CT images to detect tumors in lymphoma patients~\cite{yin2021automated}. Shwetha et al.\cite{shwetha2023yolo} performed a two-stage approach for the detection of pus cells and epithelial cells. Similarly, Kong and Shen\cite{kong2023microorganism} proposed an enhanced version of YOLO for the segmentation of microscopic images to detect microorganisms. Another study applied a combined model of YOLO and U-Net for the segmentation of microscopic images to characterize the neuron~\cite{puerta2023neuron}.\\
Furthermore, SSD and other segmentation models have been used for the detection of benign and malignant breast cancer in ultrasound images~\cite{cao2017breast}. Similarly, SSD has also been applied for the detection of various lung abnormalities in ultrasound images~\cite{kulhare2018ultrasound}, and for the detection of polyps in gastrointestinal endoscopic images using segmentation~\cite{chen2021single}. In a hybrid approach, Xu et al.~\cite{xu2023dynamic} proposed DAD-SSD (Dynamic Attention Deconvolutional Single Shot Detector) for the classification of different types of polyps in colonoscopic images. MP-FSSD (Multi-scale Pyramidal Fusion Single Shot Multibox Detector Network)~\cite{souaidi2022new} is recently proposed as a modified form of single shot detector which has been used for for the segmentation of endoscopic images frames or the endoscopy to detect polyp. \\
Similarly, different skin  diseases namely melanoma, dysplastic nevus, and healthy skin have been identified with RetinaNet~\cite{lucena2020detection}. Deep RetinaNet has been used for the segmentation of dermoscopic images to identify the melanoma lesion~\cite{rehman2022deep}. A hybrid model of RetinaNet and ResNet was applied for the detection of skin cancer in dermoscopic images~\cite{ahmed2023new}. Furthermore, Bagheri et al. ~\cite{bagheri2022skin} proposed three different object detection models for the identification of skin lesions in the dermoscopic images. 
 The detailed descriptions of these models along with the types of imaging modalities, the types of diseases these models have been used for the detection, the performance evaluation metrices and quantitative measures are shown in~\Cref{tab6:one_stage_detectors}.
\begin{table*}[htbp]
\centering
\caption{One Stage Detectors for Biomedical Imaging Segmentation}
\resizebox{1.0\linewidth}{!}{ 
\begin{tabular}{|c|c|p{2cm}|p{2.5cm}|p{3.5cm}|p{4.95cm}|p{5.0cm}|}
\hline
\textbf{Ref} & \textbf{Year} & \textbf{Modality} & \textbf{Model} & \textbf{Disease/Organ} & \textbf{Evaluation Metrics} & \textbf{Results} \\
\hline
A065~~\cite{kavitha2023brain} & 2023 & MRI & YOLO & Tumour & Specificity, Sensitivity, Accuracy & Jaccard 0.987, 0.952, 0.943, 0.955 \\
\hline
A066~~\cite{chang2018automatic} & 2018 & MRI & YOLO & Heart Disease & Dice, Hausdorff, Sensitivity & 0.9193, 10.452, 0.9085 \\
\hline
A067~~\cite{suryawanshi2024brain} & 2024 & MRI & YOLOV5 & Brain Tumour & Loss, Accuracy & 0.03, 0.9832 \\
\hline
A068~~\cite{moreno2019combined} & 2019 & MRI & YOLO+CNN & Heart Disease & PGC, Dice, APD & 98.59+-4.28\%, 0.93+-0.06, 0.72+-0.62mm \\
\hline
A069~~\cite{zhong2019real} & 2019 & MRI & YOLOV3 & Herniated Intervertebral Disc & Loss, Execution Time & 0.75-0.15, 23.9ms, 24.5ms, 24.3ms \\
\hline
A070~~\cite{unver2019skin} & 2019 & Dermoscopic Images & YOLOV3 & Skin Disease & IoU, Accuracy, Sensitivity, Specificity, Jaccard, Dice & 90, 92.99\%, 83.63\%, 94.02\%, 79.54\%, 88.13\% \\
\hline
A071~~\cite{lemay2019kidney} & 2019 & CT & YOLOV3 & Kidney Disease & Dice, IoU & 0.851, 0.759 \\
\hline
A072~~\cite{pang2019novel} & 2019 & CT & YOLOV3 & Cholelithiasis gallstones & Average Accuracy & 92.7\% for Stone, 86.50 for Cholelithiasis \\
\hline
A073~~\cite{wang2018soyolo} & 2018 & Microscopic Images & YOLOV2 & WBC Detection & Precision Rate, Recall Rate & 1.0, 0.97 \\
\hline
A074~~\cite{afshari2018automatic} & 2018 & PET & YOLO & Active Organs & Precision, Recall, ABBLE, WLE, IoU & 75-98\%, 94-100\%, 14mm, 24mm, 72\% \\
\hline
A075~~\cite{nguyen2023anomalies} & 2023 & X-Ray & YOLOV5 & Chest Disease & AP@.5, AP@.95 & 0.616, 0.322 \\
\hline
A076~~\cite{liu2023yolo} & 2023 & X-Ray & YOLOV8 & Coronary Artery & F1 Score & 0.441 \\
\hline
A077~~\cite{wu2023one} & 2023 & Angiograms & YOLOV5 & Lesion & Precision, Recall, mAP@0.1, mAP@0.5 & 0.64,0.68, 0.66, 0.49 \\
\hline
A078~~\cite{xu2023automatic} & 2023 & CT Angiograms & YOLOV4 & Pulmonary Embolism & Average Precision & 83.04\% \\
\hline
A079~~\cite{mortada2023segmentation} & 2023 & Angiograms & YOLOV7 & Heart 3 Parts & DSC & 92.63\%,85.59\%,87.57\% \\
\hline
A080~~\cite{pokhrel2023data} & 2023 & Angiograms & YOLOV8 & Heart Disease & F1 Score,mAP & 0.35,0.59 \\
\hline
A081~~\cite{osama2023empowering} & 2023 & Angiograms & YOLOV8 & Coronary Artery Stenosis & Precision,Recall,mAP & 82\%,58\%,65\% \\
\hline
A082~~\cite{zhuang2021automatic} & 2021 & Angiograms & YOLOV3 & Heart Disease & Speed,Dice,MAD,HD & 2.1-2.25 fps,93.57\%, 2.57mm,6.68mm \\
\hline
A083~~\cite{zhang2021automated} & 2021 & Ultrasound & YOLOV3 & Thyroid Nodule & Mean Precision,Mean Recall & 94.53\%,95.00\% \\
\hline
A084~~\cite{lu2022yolox} & 2022 & Ultrasound & YOLOX & Fetus Analysis & AP,SD & 0.835,0.146 \\
\hline
A085~~\cite{hussein2023myositis} & 2023 & Ultrasound & YOLOV5+YOLO-CSE & Mytosis & Accuracy,AP & 98\%,96\% \\
\hline
A086~~\cite{xie2022endoscopic} & 2022 & Ultrasound & YOLOV4 & Organs Detection & AP,mAP & 95.2\%,91.59\% \\
\hline
A087~~\cite{ghose2023improved} & 2023 & Colonoscopy & YOLOV5 & Polyp & Precision,Recall,F1-Score,mAP@0.5 & 97.26\%, 96.15\%, 95.67\%, 99.08\%, 89.44\% \\
\hline
A088~~\cite{doniyorjon2022improved} & 2022 & Endoscopy & YOLOV4 & Polyp & Accuracy & 89.9\% Training, 85.5\% Testing \\
\hline
A089~~\cite{kirthika2021yolov4} & 2021 & Endoscopy & YOLOV4 & Artefact & mAP,Speed & 49.82\%,76ms \\
\hline
A090~~\cite{durak2021deep} & 2021 & Endoscopy & YOLOV4 YOLOV3 SSD & Polyp & mAP & 86.39\%,80.98\%,55.06\% \\
\hline
A091~~\cite{alzorgani2022deep} & 2022 & Histopathology & YOLOV2 & Breast Cancer & F1 Score & 0.839 \\
\hline
A092~~\cite{rong2023deep} & 2023 & Microscopic Images & YOLO & Nuclei Tumour & Accuracy,Precision,Recall,F1 Score mIoU & 0.7110,0.8308,0.6743,0.7409 0.8423 \\
\hline
A093~~\cite{ayaz2021segmentation} & 2021 & Histopathology & YOLO & Nucleus Segmentation & F1 Score,mAP,mAR & 0.954,0.954,0.954 \\
\hline
A094~~\cite{narotamo2019segmentation} & 2019 & Microscopic Images & YOLO & Nucleus & F1 Score,Training Time & 0.98,1.6 Sec \\
\hline
A095~~\cite{nair2021mitotic} & 2021 & Histopathology & YOLOV3 & Breast Cancer & F Measure & 0.73 \\
\hline
A096~~\cite{bal2021yolo} & 2021 & Histopathology & YOLO & Breast & Accuracy, Precision, Recall, Specificity, FPR, FNR & 0.9573, 1, 0.9239, 0.0761, 0, 0 \\
        \hline
A097~~\cite{pal2020detection} & 2020 & Fundus & YOLOV3 & Lesions & Average Precision & 83.3\% \\
        \hline
A098~~\cite{lalitha2021detection} & 2021 & Fundus & YOLO-RF & Diabetic Retinopathy & Accuracy, Precision, Recall & 99.33\%, 97.20\%, 99.10\% \\
        \hline
A099~~\cite{santos2022new} & 2022 & Fundus & YOLOV5 & Lesion & mAP, F1 Score & 0.1540, 0.2521 \\
        \hline
A100~~\cite{saha2023fast} & 2023 & Fundus & YOLO + CNN & Glaucoma & F1 Score, Sensitivity, Specificity, AUC & 97.4\%, 97.3\%, 97.5\%, 99.3\% \\
        \hline
A101~~\cite{ramachandran2020fully} & 2020 & Fundus & YOLOV3 & Retinopathy & Accuracy & 100\% \\
        \hline
A102~~\cite{bagheri2020two} & 2020 & Dermoscopy & YOLO & Skin Lesion & Sensitivity, Specificity, Jaccard, bDice, Accuracy & 95.1\%, 99.5\%, 99.20\%, 92.30\%, 85.92\% \\
        \hline
A103~~\cite{singh2023fuzzy} & 2023 & Dermoscopy & YOLO & Cancer & Accuracy, Sensitivity, Specificity, Dice, Jaccard & 95.16\%, 90.86\%, 97.27\%, 86.06\%, 92.51\% \\
        \hline
A104~~\cite{yin2021automated} & 2021 & PET & YOLOV3 & Tumour & Mean Performance & 35\% \\
        \hline
A105~~\cite{shwetha2023yolo} & 2023 & Microscopic Images & YOLO & Pus Cell Detection & mAP, Accuracy & 0.87, 94.5\% \\
        \hline
A106~~\cite{kong2023microorganism} & 2023 & Microscopic Images & YOLO & Microorganism Detection & Precision, Recall, mAP@0.5 & 92.3\%, 92.8\%, 93.7\% \\
        \hline
A107~~\cite{puerta2023neuron} & 2023 & Microscopic Images & YOLO + U-Net & Neurons Characterization & MACC, MLO, MIoU, SDACC, SDLO, SDIoU & 0.9714, 0.0755, 0.8026, 0.0224, 0.0542, 0.1593 \\
        \hline
A108~~\cite{cao2017breast} & 2017 & Ultrasound & SSD & Breast Cancer & APR, ARR, F1 Score & 97.20\%, 70.56\%, 81.76\% \\
        \hline
A109~~\cite{kulhare2018ultrasound} & 2018 & Ultrasound & SSD & Lungs Diseases & Sensitivity, Specificity & 93.6\%, 96.5\% \\
        \hline
A110~~\cite{souaidi2022new} & 2022 & Endoscopy & SSD & Polyp & mAP & 92.67\% \\
        \hline
A111~~\cite{chen2021single} & 2021 & Endoscopy & SSD & Polyp & mAP & 95.74\% \\
        \hline
A112~~\cite{xu2023dynamic} & 2023 & Colonoscopy & SSD & Polyp & mAP, Accuracy & 76.55\%, 74.40\% \\
        \hline
A113~~\cite{lucena2020detection} & 2020 & Dermoscopy & RetinaNet & Melanoma Dysplastic Detection & Accuracy & 68.8\%, 72.5\% \\
        \hline
A114~~\cite{rehman2022deep} & 2022 & Dermoscopy & RetinaNet & Melanoma & Average Precision & 97\% \\
        \hline
A115~~\cite{ahmed2023new} & 2023 & Dermoscopy & RetinaNet + ResNet & Skin Cancer & Jaccard, Dice, Sensitivity, Specificity, AUC & 91.4\%, 90.70\%, 91.60\%, 97.80\%, 97.60\% \\
        \hline
A116~~\cite{bagheri2022skin} & 2022 & Dermoscopy & RetinaNet & Skin Lesion & Sensitivity, Specificity, Jaccard, Dice, Accuracy & 88.56\%, 96.25\%, 80.02\%, 87.62\%, 94.37\% \\
        \hline
\end{tabular}
}
\label{tab6:one_stage_detectors}
\end{table*}
\subsubsection{Analysis of One Stage Detectors}
Similar to two-stage detection models, the modalities targeted by one-stage detection models include MRI, CT, PET, X-rays, dermoscopic images, microscopic images, ultrasound, endoscopic, colonoscopic, histopathology, fundus images, and several other modalities. Among the one-stage detection models, YOLO and its various versions are the most powerful and frequently applied. The most commonly used YOLO version for medical image segmentation is YOLOv3, followed by YOLOv5, YOLOv4, and YOLOv8, respectively. Some authors have also applied YOLOv2 and YOLOv7, while others have used SSD, RetinaNet, and RefineNet for biomedical image segmentation, though their frequency of use is much lower.
The quantitative measures of different performance evaluation parameters are shown in the last column of the table, where considerable variations can be observed. These variations result from several factors, including ease of processing and implementation, lower computational cost, high accuracy, simple architecture, and robustness in handling complex datasets. These factors contribute to YOLO and its versions outperforming other one-stage models.
\subsection{Articles Related to Point-Based Detection Models}
Some object detection models for image segmentation possess properties distinct from both one-stage and two-stage detection models. Due to these unique characteristics, they have been categorized separately as advanced object detection models or point-based detection models. The most common models in this category include CornerNet, CornerNet-Lite, Objects as Points, and FoveaBox. In the literature, many authors have applied these models to the segmentation of biomedical images.
In this regard, CornerNet has been used  for the segmentation of brain MRI to analyse it for brain tumour detection~\cite{nawaz2021analysis},  for the segmentation of colonoscopic images to detect polyp~\cite{wang2022afpmask}, and for  the extraction of informative features from the X-ray images to segment it for the detection of knee osteoarthritis~\cite{aladhadh2023knee}. It has also been used 
 for the segmentation of MRI images to detect two types of malignant cancers namely low-grade glioma and high-grade glioma~\cite{sandhya2023brain}.
Another study proposed a backbone of CornerNet for the segmentation of brain MRI to detection brain tumour~\cite{alevizos2023novel}. A combined model of CornerNet and Fuzzy K-means has been for the segmentation of dermoscopic images to detect melanoma in the skin~\cite{nawaz2022mseg}.   \\ 
Recently, two deep learning models namely centernet and U-net have been used for detection and segmentation of endoscopic images to detect gastrointestinal disease ~\cite{choi2020centernet}. Another recent study proposed an improved version of Centernet for the segmentation of microscopic images to detect white blood cells in them~\cite{zheng2023white}. Similarly, Jiang et al. ~\cite{jiang2023eccpolypdet} proposed a modified version of CenterNet for the segmentation of colonoscopic images to detect polyp. In other studies the CenterNet has been used for the segmentation of Optical Coherence Tomography Angiography (OCTA) images to detect retinal vascular bifurcation and crossover points~\cite{wang2021detection}, and  for the segmentation of brain MRI images to detect tumour~\cite{masood2022brain}.
\\
Similarly, Albahli and Nazir ~\cite{albahli2022aicenter} used CenterNet for the segmentation of X-Ray images to classify and localize the diseases in them. An other study applied CenterNet for the segmentation of CT images to detect pulmonary cancers~\cite{thangavel2024effective}. Jasitha and Pournami ~\cite{jasitha2023glomeruli} analysed both the faster R-CNN and CenterNet for the segmentation of histopathology images to detect Glomeruli. Recently, Nazir et al. ~\cite{nazir2021detection} applied CenterNet for the segmentation of retinal images to detect diabetic retinopathy. In hybrid models, the combination of R-CNN, CenterNet, and deep snake for thyroid nodule detection in the ultrasound images~\cite{shen2021cascaded}. Another hybrid model of VFNet, FoveaBox, FCOS, and DetectoRS for detection of lymph nodes in MRI~\cite{mathai2021detection}.
Mattikalli et al. ~\cite{mattikalli2022universal} developed an ensemble model of VFNet, FoveaBox, and RetinaNet for the detection of lesion in CT images.  \\
Sakunpaisanwari et al. ~\cite{sakunpaisanwari2022blood} different two stage detector models and one stage detector models including CornerNet and CenterNet for blood vessels detection in CT images. Another study implemented different two stage detectors, one stage detectors, and other advanced detection models including CenterNet, for the detection of gastric polyps in the endoscopic images~\cite{durak2021deep}. Recently, an ensembled model of foveabox with other one stage detectors have been used to detect lymphadenopathy in MRI images~\cite{mathai2023universal}. Similarly, Celik et al. ~\cite{celik2023role} studied the role of different deep learning models including foveabox in the detection of lesions on panoramic radiographs. An other article~\cite{cho2021automatic} compared the performance of RetinaNet with cornernet and centernet in automatic tip detection of surgical instruments in biportal endoscopic spine surgery. Similarly, in an other study~\cite{sakunpaisanwari2022blood}, different deep learning models including centernet and cornernet are used for blood vessels detection in CT scans of lower extremities.\\
Nguyen et al. ~\cite{nguyen2021circle} applied different deep learning models including centernet and cornernet for the detection of glomeruli and nuclei in pathological images. Furthermore, Zheng et al. ~\cite{zheng2023white} applied centernet for the detection of blood cells in the microscopic images. In another work, a centernet based on heatmap pyramid structure for the detection of rib fracture in CT images of chest~\cite{su2023rib}. Faster R-CNN, RetinaNet, and Centernet have been used for the detection and characterization of parasite eggs in the microscopic images~\cite{kitvimonrat2020automatic}. Similarly, Jiang et al. ~\cite{jiang2022feature} used centernet along with Resnet 50 for feature extraction in the diagnosis of ultrasound images. In another study use of YOLOv4 and Centernet for the diagnosis of breast ultrasound to detect lesion~\cite{dai2021more}. Recently, few deep learning models including centernet have been used for the detection of polyp segmentation in colonoscopy~\cite{wang2022afpmask}. Transformer based centernet is also introduced which is used for the liver tumor detection in CT images~\cite{ma2024transformer}.
The detailed descriptions of all these models along with other specifications are shown in~\Cref{tab7:advanced_detectors}.
\begin{table*}[htbp]
    \centering
    \caption{Advanced or point-based detectors for biomedical imaging segmentation.}
    \label{tab7:advanced_detectors}
\resizebox{1.0\linewidth}{!}{ 
\begin{tabular}{|c|c|p{2cm}|p{3.5cm}|p{3cm}|p{4.5cm}|p{4cm}|}
        \hline
        \textbf{ID-Ref} & \textbf{Year} & \textbf{Modality} & \textbf{Model} & \textbf{Disease/Organ} & \textbf{Evaluation Metrics} & \textbf{Results} \\
        \hline
        A117~~\cite{nawaz2021analysis} & 2021 & MRI & CornerNet & Brain Tumour & Accuracy, Precision, Recall & 98.7\%, 97.4\%, 96.9\% \\
        \hline
        A118~~\cite{nawaz2022mseg} & 2022 & Dermoscopic Images & CornerNet & Skin Disease & mAP, Sensitivity, Specificity, Accuracy & 99.48\%, 99.39\%, 99.63\%, 98.8\% \\
        \hline
        A119~~\cite{wang2022afpmask} & 2022 & Colonoscopic Images & CornerNet & Polyp & Precision, Recall & 99.36\%, 96.4\% \\
        \hline
        A120~~\cite{alevizos2023novel} & 2023 & MRI & CornerNet & Brain Tumour & Accuracy & 89.8\% \\
        \hline
        A121~~\cite{aladhadh2023knee} & 2023 & X-Ray & CenterNet & Knee Osteoarthritis & Accuracy & 99.14\% \\
        \hline
        A122~~\cite{sandhya2023brain} & 2022 & MRI & CornerNet & Cancer & Accuracy, Precision, Recall & 91.3\%, 91\%, 88\% \\
        \hline
        A123~~\cite{choi2020centernet} & 2020 & Endoscopic Images & CenterNet, U-Net & Gastrointestinal Disease & mAP & 0.1932 \\
        \hline
        A124~~\cite{zheng2023white} & 2023 & Microscopic Images & CenterNet & WBC Detection & mAP & 98.8\% \\
        \hline
        A125~~\cite{jiang2023eccpolypdet} & 2023 & Colonoscopic Images & CenterNet & Polyp & Precision, Recall, F1-Score, Average Precision & 87.7\%, 84.2\%, 85.8\%, 82.2\% \\
        \hline
        A126~~\cite{wang2021detection} & 2021 & OCTA & CenterNet & Retinal Vascular & Average Precision, Crossover Points, mAP & 80.81\%, 85.86\%, 83.34\% \\
        \hline
        A127~~\cite{masood2022brain} & 2022 & MRI & CenterNet & Brain Tumour & Precision, Recall, F1 Score, Accuracy & 99.52\%, 99.92\%, 99.72\%, 98.07\% \\
        \hline
        A128~~\cite{albahli2022aicenter} & 2022 & X-Ray & CenterNet & Different Organs & AUC, Average IoU & 0.888, 0.801 \\
        \hline
        A129~~\cite{thangavel2024effective} & 2024 & CT & CenterNet & Pulmonary Cancer & DSC, Sensitivity, PPV & 99.07\%, 99.65\%, 98.96\% \\
        \hline
        A130~~\cite{jasitha2023glomeruli} & 2023 & Histopathology Images & CenterNet & Glomeruli & IoU, mAP & 64.2\%, 65.7\% \\
        \hline
        A131~~\cite{nazir2021detection} & 2021 & Fundus & CenterNet & Diabetic Retinopathy & Average Accuracy & 98.10\% \\
        \hline
        A132~~\cite{shen2021cascaded} & 2021 & Ultrasound & CenterNet & Thyroid Nodule & Accuracy, Mean IoU & 77.92\%, 79.58\% \\
        \hline
        A133~~\cite{mattikalli2022universal} & 2022 & CT & FoveaBox & Lesion & mAP & 50.51\% \\
        \hline
        A134~~\cite{mathai2021detection} & 2021 & MRI & FoveaBox & Lymph Nodes & mAP & 61.67\% \\
        \hline
        A135~~\cite{sakunpaisanwari2022blood} & 2022 & CT & CornerNet, CenterNet & Blood vessels & Precision, Recall & 0.847, 0.865, 0.939, 0.703 \\
        \hline
        A136~~\cite{durak2021deep} & 2021 & Endoscopic & CenterNet & Gastric Polyps & mAP, IoU & 60.02\%, 58.27\% \\
        \hline
        A137~~\cite{mathai2023universal} & 2023 & MRI & FoveaBox & Lymphadenopathy & mAP & 52.3\% \\
        \hline
        A138~~\cite{celik2023role} & 2023 & Radiographs & Foveabox & Lesion & mAP & 0.726 \\
        \hline
        A139~~\cite{cho2021automatic} & 2021 & Endoscopic & CornerNet, CenterNet & Spine surgery & Recall, Precision & 0.796, 0.877, 0.839, 0.807 \\
        \hline
        A140~~\cite{sakunpaisanwari2022blood} & 2022 & CT & CornerNet, CenterNet & Lower Extremities & Precision & 0.847, 0.939 \\
        \hline
        A141~~\cite{nguyen2021circle} & 2021 & Pahtological & CornerNet, CenterNet & Glomeruli & AP & 0.595, 0.574 \\
        \hline
        A142~~\cite{zheng2023white} & 2023 & Microscopic & CenterNet & Blood Cells & F1, AP, mAP & 0.854, 0.934, 0.788 \\
        \hline
        A143~~\cite{su2023rib} & 2023 & CT & CenterNet & Rib fracture & mAP & 89.2\% \\
        \hline
        A144~~\cite{kitvimonrat2020automatic} & 2020 & Microscopic & CenterNet & Parasite eggs & mAP, Weighted mAP & 0.5000, 0.2612 \\
        \hline
        A145~~\cite{jiang2022feature} & 2022 & Ultrasound & CenterNet & Clinical Diagnosis & mAP & 78.6\% \\
        \hline
        A146~~\cite{dai2021more} & 2021 & Ultrasound & CenterNet & Breast cancer & mAP & 13.8\% \\
        \hline
        A147~~\cite{wang2022afpmask} & 2022 & Colonoscopy & CenterNet & Polyp detection & Precision, Recall, F1 Score & 95.27\%, 93.35\%, 94.30\% \\
        \hline
        A148~~\cite{ma2023automatic} & 2024 & CT & CenterNet & Liver tumour & Recall, Precision, mAP & 0.8279, 0.8343, 0.8489 \\
        \hline
    \end{tabular}
    }
\end{table*}

\subsubsection{Analysis of Point-Based Detectors}
Advanced (or Point-Based) detectors have been applied for the segmentation of various biomedical imaging modalities, including MRI, X-rays, dermoscopic images, colonoscopic images, microscopic images, OCTA, CT images, histopathological images, fundus images, ultrasound images, endoscopic images, and radiographs. When their performance is closely analyzed, the highest accuracies have been achieved using both CornerNet and CenterNet, with similar results in some cases and slight fluctuations between them. In certain instances, CornerNet outperforms CenterNet, while in others, CenterNet surpasses CornerNet. The results obtained from FoveaBox, however, are less satisfactory in terms of accuracy. The variations in these outcomes can be attributed to several factors, including the size of datasets used in the experiments, the running environment, model parameter tuning, data division into training and testing sets, data augmentation approaches, and data pre-processing and post-processing methods.

%% file: content/06-FutureDirections.tex
\section{CHALLENGES \& FUTURE DIRECTIONS}
\label{sec:futureDir}
There are many challenges associated with different types of biomedical images as well the models applied for their segmentation. This section outlines all the challenges suggest future directions related to these challenges and future trends in biomedical image segmentation used for diseases detection. 
\subsection{Challenges with Biomedical Image Segmentation}
 Biomedical image segmentation faces many challenges including the complex nature of image data, variability in images, high computational requirements, insufficient data availability, and differing hardware, software requirements, 
 an many other factors. The foremost challenge in this regard is the accurate annotation and labeling process. Unlike classification, which only requires a few output values, segmentation needs generating images as output for preparing training datasets. This task is not possible without the involvement of field experts and high-quality software tools, which require additional time, effort, and cost. Additionally, the complex structure of biomedical images—encompassing various shapes, formats, organ anatomies, cell structures, and tissue architectures—significantly impacts the performance of segmentation models. Another major challenge is the scale variation in biomedical images, ranging from the small nuclei of cells to large full organs like the heart, lungs, kidneys, and others. Developing robust segmentation models to accommodate these large-scale differences is particularly challenging. \\
Similarly, another important and considerable issue related to biomedical images is the class imbalance problem ~\cite{ma2023automatic}. This issue arises due to unavailability of sufficient instances for all the available classes of diseases being considered in the problem. This problem has a very bad impact on the performance of the segmentation models which leads to higher inaccuracy in the segmentation process and biases in the prediction. Similarly, all types of biomedical images are affected by different types of noises which adequately degrade the quality of images. Other factors affecting the quality of medical images include low contrast, distortions, low quality illumination, and body movements during .capturing images. All these types of unwanted activities have negative impact on biomedical images processing resulting in poor performance of the segmentation tasks. To address all of these challenges, we need effective, efficient, and powerful communication and collaborations among different experts including medical experts, computer experts, and scientists. This will reduce the barriers prevailing in the processing of biomedical images which will improve the process of segmentation.  
\subsection{Challenges with Segmentation Models}
There are different challenges associated with two stage detection models, one stage detection models, and the advanced detection models. All these challenges are highlighted in this section. The most important issue associated with two stage detection models is the complex architecture of the models due to performing the segmentation task in two stages namely region proposals and the classification. The second challenge with these models is their complete dependency on the mechanism of region proposal for generating bounding boxes which may result in the localization error which significantly decrease the segmentation accuracy of the models. Another main problem with the two stage detection models is the slow processing due to involvement of multiple stages for the segmentation task. Similarly, the models may be poorly trained due to optimization of two different components namely region proposals generation and the classification. To improve the convergence and ensuring the effectiveness in the learning both the training mechanism and the hyperparameter tuning process need to be highly careful and managed. \\
The major challenge of one stage detection model for biomedical image segmentation is low accuracy. The low accuracy results from the fact that these models directly perform the segmentation process with region proposals. Another major issue with one stage detectin model is their weakness in segmentation of small regions showing the abnormality. \\
Since, there is no specific mechanism for the identirication of different regions, therefore, it is very difficult for the detection to distinguish between the background and foreground of the biomedical images. Likewise, identification of different sizes, shapes, and textrure of various abnormal existence in the biomedical images is another challenge for one strage detection models. Another difficult mechanism for one stage detector is collecting contextual information from the surroundings of the affected areas which degrades the performance of the model in the segmentation process.\\ 

\subsection{Future Research Directions}
This review paper covers various object detection-based models for biomedical imaging segmentation. Although these models represent the latest architectures for biomedical image segmentation, the complexity of the segmentation process still leaves considerable room for improvement. This section outlines potential future research directions in biomedical imaging segmentation.\\
\textbf{1) Multi-modal Approach:}
Current architectures for biomedical imaging segmentation typically rely on a uni-modal approach, where a model is used to segment images from a single modality. A significant direction for researchers is the application of a multi-modal approach, integrating information from multiple biomedical imaging modalities such as CT, X-Ray, MRI, and SPECT. Combining information from various modalities could greatly enhance the performance and accuracy of the segmentation process.\\
\textbf{2) Transfer Learning:}
A major challenge in biomedical imaging segmentation is the scarcity of sufficient datasets, which complicates the segmentation process. This issue could be addressed through transfer learning, where models trained on large datasets are adapted for use on smaller datasets. This approach could provide a solid foundation for researchers and industries focused on biomedical imaging segmentation.\\
\textbf{3) Advanced Deep Learning Models:}
While the deep learning models discussed in this review have contributed significantly to biomedical imaging segmentation, there is still potential for improvement. Advanced models, such as transformer-based architectures, Graph Neural Networks (GNN), hybrid models, and attention-based models, could further enhance segmentation accuracy, especially in images compromised by various types of noise.\\
\textbf{4) Unsupervised and Semi-Supervised Approaches:}
Labeling and annotating biomedical images require significant resources, including time and effort. Unsupervised and semi-supervised learning approaches could alleviate some of these challenges. Models such as Self Organizing Maps (SOM), Generative Adversarial Networks (GANs), graph-based techniques, and auto-encoders represent promising areas for future research.\\
\textbf{5) Enhancement Techniques:}
Biomedical images often suffer from quality issues such as noise, poor illumination, and suboptimal image capture methods. Research into models and techniques that enhance image quality could improve segmentation accuracy and performance.\\
\textbf{6) Clinical Aspects:}
Bridging the gap between technological advancements and clinical applicability is crucial. Future research should focus on developing methods that support the practical implementation of segmentation models, aiding in the detection of abnormalities across various human organs.\\
\textbf{7) Hybrid approaches with large data:}
Exploring interdisciplinary approaches that combine biomedical imaging the data from other related domains such as  bio-informatics, computational physics, and systems biology could also be explored to analyse the images based on different types of related data. For example, with the use of genetic data and integrating it in image segmentation  could enhance the understanding of disease mechanisms at a molecular level, leading to more personalized and effective treatments.\\
\textbf{8) Real-time Segmentation:}
Developing methods for real-time segmentation of biomedical images could significantly impact clinical diagnostics and surgical procedures. This would involve optimizing existing algorithms for speed without sacrificing accuracy, potentially utilizing edge computing technologies to process data directly on medical devices and providing instant results for clinical decision-making.\\
\textbf{9) Interactive Segmentation:} Enhancing segmentation results through user interaction is a promising avenue. Exploring the latest Large Language Models (LLMs) could provide new methods for user interaction with segmentation systems, potentially making them more intuitive and effective for various medical imaging tasks.

%% file: content/07-Conclusion.tex
\section{CONCLUSION}
\label{Sec:conc}
We conducted a comprehensive review of the deep learning based object detection methods use for  biomedical image segmentation. Following, the standard SRL protocol, we selected 148 articles, and extracted information from these auricles. The extracted information include deep learning models, biomedical imaging modalities, and human disease. We have  presented these information in tabular forms and provided various charts and classifications.  Object detection based deep learning models are divided into three types namely two stage detectors, one-stage detectors, and advanced detectors which have been reviewed, critically analyzed. Finally, we highlighted different challenges associated with medical imaging modalities and object detection based deep learning models used for their segmentation, and highlighted future directions.  To the best of our knowledge, this is the first SLR conducted on this specific topic within the field. We believe this work will serve as a foundational reference, providing essential guidance for new researchers and offering a comprehensive overview for both new and experienced researchers in the field. 